\documentclass[3p,times,11pt]{elsarticle}
\usepackage{algorithm,algorithmic}
\usepackage[titletoc,title]{appendix}

\usepackage{mdframed}
\usepackage{float}
\usepackage{wrapfig,blindtext}
\usepackage{graphicx} 
\usepackage{epstopdf} 
\usepackage{pslatex} 
\usepackage{amsmath,amssymb}
\usepackage{caption}
\usepackage{subcaption}
\usepackage{amsfonts,amsthm} 
\usepackage[english]{babel} 
\usepackage[dvipsnames]{xcolor}
\usepackage{enumitem}
\usepackage{booktabs}
\usepackage{wasysym}
\usepackage{hyperref}
\usepackage{pdfpages}
\usepackage{float}
\usepackage{comment}
\usepackage{soul}
\newtheorem{theorem}{Theorem}

\usepackage{nomencl}
\usepackage{multirow}
\usepackage{soul}
\usepackage{cancel}
\usepackage[utf8]{inputenc}
\usepackage[english]{babel}

\usepackage[font=small,labelfont=bf,
   justification=justified,format=plain]{caption}

\usepackage{comment}

\usepackage{nomencl}
  \makenomenclature

\begin{document}
\topmargin -1.5cm
\textheight 23cm
\begin{frontmatter}

\title{Disentangling Generative Factors of Physical Fields  Using Variational Autoencoders}

\author{Christian Jacobsen}
\ead{csjacobs@umich.edu}
\author{Karthik Duraisamy}
\ead{kdur@umich.edu}
\address{University of Michigan, Ann Arbor, MI}
\newpage

\begin{abstract}\label{sec:abstract}
The ability to extract generative parameters from high-dimensional fields of data in an unsupervised manner is a highly desirable yet unrealized goal in computational physics. This work explores the use of variational autoencoders (VAEs) for non-linear dimension reduction with the specific aim of {\em disentangling} the low-dimensional latent variables to identify independent physical parameters that generated the data. 
A disentangled decomposition is interpretable, and can be transferred to a variety of tasks including generative modeling, design optimization, and probabilistic reduced order modelling.  A major emphasis of this work is to characterize disentanglement using VAEs while minimally modifying the classic VAE loss function (i.e. the Evidence Lower Bound) to maintain high reconstruction accuracy.  The loss landscape is characterized by over-regularized local minima  which  surround  desirable solutions. We illustrate comparisons between disentangled and entangled representations by juxtaposing learned latent distributions and the true generative factors in a model porous flow problem.  Hierarchical priors are shown to  facilitate the learning of disentangled representations.  The regularization  loss is  unaffected  by  latent  rotation  when  training  with rotationally-invariant priors, and thus learning non-rotationally-invariant priors aids  in capturing the properties of generative factors, improving disentanglement. 
Finally, it is shown that semi-supervised learning - accomplished by labeling a small number of samples ($O(1\%)$) - results in accurate disentangled latent representations that can be consistently learned.
\end{abstract}

\end{frontmatter}

\section{Introduction} \label{sec:intro}
Unsupervised representation learning is a popular area of research because of the need for low-dimensional representations in unlabeled data. 
Low-dimensional {\em latent} representations of high- dimensional data have many applications ranging from facial image generation \cite{8545633} and music generation \cite{9262797} to autonomous controls \cite{inproceedings} among many others. Generative adversarial networks (GANs) \cite{goodfellow2014generative}, variational autoencoders (VAEs) \cite{kingma2014autoencoding} and their variants \cite{Higgins2017betaVAELB, klushyn2019learning, kim2019disentangling, zhao2018infovae}, among other methods,  aim to approximate an underlying distribution $p(y)$ of high-dimensional data through a two-step process.  Compressed representations $z$ are sampled from a low-dimensional - yet unknown - distribution $p(z)$. In the case of VAEs, which is the focus of this work, an encoding distribution $p(z|y)$ and a decoding distribution are learned simultaneously by maximizing a bound on the likelihood of the data (i.e. the evidence lower bound (ELBO) \cite{kingma2014autoencoding}). Thus, a mapping from the high-dimensional space to a low-dimensional space and the corresponding inverse mapping is learned simultaneously, allowing approximations of both $p(y)$ \emph{and} $p(z)$. Learning the lower-dimensional representation, or \emph{latent space}, can facilitate computationally-efficient data generation and extract only the information necessary to reconstruct the data \cite{bengio2014representation}. Modifications to the ELBO objective have been suggested in the literature, primarily with improved disentanglement in mind. The $\beta$-VAE \cite{Higgins2017betaVAELB} was introduced to improve disentanglement by adjusting the weight of regularization loss. FactorVAE \cite{kim2019disentangling} introduces a total correlation term (TC) to encourage learning a factorized latent representation. InfoVAE \cite{zhao2018infovae} augments the ELBO with a term to promote maximization of the mutual information between the data and the learned representation. Many other developments based on the VAE objective have been introduced in the literature. 
VAEs have been implemented in many applications including inverse problems \cite{tait2020variational}, extracting physical parameters from spatio-temporal data \cite{Lu_2020}, and constructing probabilistic reduced order models \cite{lopez, 2021pdedriven}, among others. 

To illustrate the idea of disentanglement and its implications, consider a dataset consisting of images of teapots \cite{eastwood2018a}. Each image is generated from 3 parameters indicating the color of the teapot (RGB) and 2 parameters corresponding to the angle the teapot is viewed from. Thus, even though the RGB image may be very high dimensional, the intrinsic dimensionality is just 5. Representation learning can be used to extract a low-dimensional latent model containing useful and meaningful representations of the high-dimensional images. Learned latent representations need not be \emph{disentangled} to be useful in some sense, but disentanglement enhances interpretability of the representation. Disentanglement references a structure of the latent distribution in which changes in each parameter in the learned representation correspond directly to changes in a single yet different generative parameter. In an unsupervised setting, one cannot guarantee that a disentangled representation can be learned. 

The requirement for disentanglement depends on the task at hand, but a disentangled representation may be used in many tasks containing different objectives. Indeed, Bengio et al.~\cite{bengio2014representation} state that  `the most robust approach to feature learning is to disentangle as many factors as possible, discarding as little information about the data as is practical'. In the teapot example, changes in one of the learned latent dimensions may correspond to changes in the color red and one of the viewing angles, which would indicate an entangled representation.  Another example, more relevant to our work, is that of fluid flow over an airfoil. 
Learning a disentangled representation of the flow conditions along with the shape parameters using VAEs can allow rapid prediction of the flow field with interpretability of the latent representation, facilitating efficient computation of the task at hand. The disentangled representation can be transferred to a variety of tasks easily such as design optimization, developing reduced order models in the latent space or parameter inference from flow fields. It is the ability of disentangled representations to transfer across tasks with ease and interpretability which makes them so useful. In many practical physics problems, full knowledge regarding the underlying generative parameters of high-dimensional data may not exist, thus making it challenging to ascertain the quality of disentanglement.

Disentanglement using VAEs was first addressed in the literature~\cite{Higgins2017betaVAELB,kim2019disentangling} by modifying the strength of regularization in the ELBO loss, with the penalty of sub-optimal compression and reconstruction. FactorVAEs \cite{Higgins2017betaVAELB} encourage a factorized representation, which can be useful for disentanglement in the case of independent generative parameters, but undesirable when parameters are correlated. Rolinek et al.~\cite{rolinek2019variational} suggest that the ability of the VAE to learn disentangled representations is not inherent to the framework itself, but an "accidental" byproduct of the typically assumed factorized form of the encoder. The prior distribution is of particular importance as the standard normal prior often assumed allows for rotation of the latent space with no effect on the ELBO loss. Disentangled representations are still often learned due to a factorized form of the encoding distribution with sufficiently large weight on regularization. Additional interpretations and insight into the disentanglement ability of VAEs are found in \cite{burgess2018understanding}.

Our work on unsupervised representation learning is motivated from a computational-physics perspective. We focus on the application of VAEs for use with data generated by partial differential equations (PDEs). The central questions we seek to answer in this work are: a) can we reliably disentangle parameters from data obtained from PDEs governing physical problems using VAEs, and b) what are the characteristics of disentangled representations? Learning disentangled representations can be useful in many capacities: developing probabilistic reduced order models, design optimization, parameter extraction, and data interpolation, among others. Many of the applications of such representations, and the ability to transfer between them, rely heavily on the disentanglement of the latent space. Differences in disentangled and entangled representations are identified, and conclusions are drawn regarding the inconsistencies in learning such representations. Our goals are not to compare the available methods to promote disentanglement, as in \cite{locatello2019challenging}, but rather to illustrate the use of VAEs without modifying the ELBO and to understand the phenomenon of disentanglement itself in this capacity. The use of hierarchical priors is shown to greatly improve the prospect of learning a disentangled representation in some cases without altering the standard VAE loss through the learning of non-rotationally-invariant priors. Along the way, we provide intuition on the objective of VAEs through connections to rate-distortion theory, illustrate some of the challenges of implementing and training VAEs, and provide potential methods to overcome some of these issues such as 'vanishing KL' \cite{4839047}. 

The outline of this paper is as follows: In Section 2, we introduce the VAE, connect it to rate-distortion (RD) theory, discuss disentanglement, and derive a bound on the classic VAE loss (the ELBO) using hierarchical priors (HP). In Section 3, we introduce a sample application of Darcy flow as the main illustrative example of this work. In Section 4, we present challenges in training VAEs and include possible solutions, and investigate the ELBO loss landscape. We illustrate disentanglement of parameters on the Darcy flow problem, and provide insight into the phenomenon of disentanglement in Section 5. In Section 6, conclusions and insights are drawn on the results of our work, and future directions are discussed.

The numerical experiments in this paper can be recreated using our code provided in \url{https://github.com/christian-jacobsen/Disentangling-Physical-Fields}.

\section{Variational Autoencoder Formulation}
\label{sec:vae}


It is desirable that the latent distribution contains only enough information about the data to allow accurate reconstruction of the data, leading to an optimally compressed representation. The VAE framework is a method of data compression with many ties to information theory \cite{Alemi2018FixingAB}. In applications with little to no knowledge regarding the nature of obtained data, the latent factors extracted using VAEs can act as a set of features describing the generative parameters underlying the data. A direct correlation between the generative parameters and the compressed representation, or a disentangled representation, is sought such that the representation can be applied to a multitude of downstream tasks. Some example tasks include performing predictions on new generative parameters, interpreting the data in the case of unknown generative parameters, and computationally efficient design optimization.

Data snapshots obtained from some physical system or a model of that system is represented here by random variable $Y: \Omega \rightarrow \mathcal{Y}$ where $\Omega$ is a sample space and $\mathcal{Y}$ is a measurable space ($\mathcal{Y} = \mathbb{R}^m$ is typically assumed). Each realization of $Y$ is generated from an function of $\Theta$ such that $\Theta : \Omega \rightarrow \mathbb{R}^p$ is a random variable representing generative parameters with distribution $p(\theta)$. With no prior knowledge of $\Theta$, the random variable $Z: \Omega \rightarrow \mathbb{R}^n$ represents the latent parameters to be inferred from the data. A probabilistic relationship between $\Theta$ and $Y$ is sought in an unsupervised manner using only samples from $p(y)$.

The VAE framework infers a latent-variable model by replacing the posterior $p(z|y)$ with a parameterized approximating posterior $q_\phi(z|y)$ \cite{kingma2014autoencoding}, known as the encoding distribution. A parameterized decoding distribution $p_\psi(y|z)$ is also constructed to predict data samples given samples from the latent space. Only the encoding distribution and the decoding distribution are learned in the VAE framework, but the  \emph{aggregated posterior} $q_\phi(z)$ (to the best of our knowledge, first referred to in this way by \cite{makhzani2016adversarial}), is of particular importance in disentanglement. It is defined as the marginal latent distribution induced by the encoder
\begin{equation} \label{eq:agg_post}
    q_\phi(z) \triangleq \int_{\mathcal{Y}} p(y) q_\phi(z|y) dy \; , 
\end{equation}
where the true data distribution is denoted by $p(y)$. The induced data distribution is the marginal output distribution induced by the decoder 
\begin{equation}\label{eq:induced_data}
    p_\psi(y) \triangleq \int_{\mathbb{R}^n} p(z)p_\psi(y|z) dz \;.
\end{equation}
It is noted that the true data distribution is typically unknown; only samples of data $\{y^{(i)}\}_{i=1}^N$ are available. The empirical data distribution is thus denoted $\hat{p}(y)$, and any expectation with respect to the empirical distribution is simply computed as an empirical average
\[
\mathbb{E}_{\hat{p}(y)}[f(y)] \triangleq \frac{1}{N}\sum_{i=1}^N f(y^{(i)}) \; . 
\]

Learning the latent model is accomplished by simultaneously learning the encoding and decoding distributions through maximizing the evidence lower bound (ELBO), which is a lower bound on the log-likelihood \cite{odaibo2019tutorial}. To derive the ELBO loss, we begin by expanding the relative entropy between the data distribution and the \emph{induced} data distribution
\[
D_{KL}[p(y)||p_\psi(y)] = \mathbb{E}_{Y\sim p(y)}[\log p(y)] - \mathbb{E}_{Y\sim p(y)}[\log p_\psi(y)]
\]
where the first term on the right hand side is the negative differential entropy of $Y$, $-H(Y)$. Noting that relative entropy is always greater than or equal to zero and introducing \[
p_\psi(y) = \frac{p_\psi(y|z)p(z)p_\phi(z|y)}{p(z|y)p_\phi(z|y)} \; ,
\]
we arrive at the following inequality
\[
H(Y) + \mathbb{E}_{Y\sim p(y)}[D_{KL}[p_\phi(z|y)||p(z|y)]] \leq \mathbb{E}_{p(y)}[\mathbb{E}_{q_\phi(z|y)}[\log p_\psi(y|z)]] -\mathbb{E}_{p(y)}[D_{KL}[q_\phi(z|y)||p(z)] \; .
\]
Thus, 
\begin{equation} \label{eq:log_like}
    \mathbb{E}_{p(y)}[\log(p(y))] \geq \mathbb{E}_{p(y)}[\mathbb{E}_{q_\phi(z|y)}[\log p_\psi(y|z)]] -\mathbb{E}_{p(y)}[D_{KL}[q_\phi(z|y)||p(z)] \; ,
\end{equation}
where $p(z)$ is a prior distribution. The prior is  specified by the user in the classic VAE framework. The right-hand side in (\ref{eq:log_like}) is the well-known ELBO. Maximizing this lower bound on the log-likelihood of the data is done by minimizing the negative ELBO. The optimization is performed by learning the encoder and decoder parameterized as neural networks. This gives the VAE loss function
\begin{equation} \label{eq:L_VAE}
    \mathcal{L}_{VAE} = \mathbb{E}_{\hat{p}(y)}[D_{KL}[q_\phi(z|y)||p(z)] +  \mathbb{E}_{\hat{p}(y)}[\mathbb{E}_{q_\phi(z|y)}[-\log p_\psi(y|z)]] \; ,
\end{equation}
where the first term on the right-hand side is the regularization loss $L_{REG}$ and drives the encoding distribution closer (in the sense of minimizing KL divergence) to the prior distribution. The second term on the right-hand side is the reconstruction error $L_{REC}$ and encourages accurate reconstruction of the data. 

Selecting the prior distribution as well as the parametric form of the encoding and decoding distribution can allow closed form solutions to compute $\mathcal{L}_{VAE}$. The prior distribution is often conveniently chosen as a standard normal distribution
\[
p(z) = \mathcal{N}(z;0, I_{n\times n}) \; .
\]
The encoding and decoding distributions are also often chosen as factorized normal distributions
\[
q_\phi(z|y) = \mathcal{N}(z; \mu_\phi(y), \textrm{diag}(\sigma_\phi(y)))
\]
and
\[
p_\psi(y|z) = \mathcal{N}(y; \mu_\psi(z), \textrm{diag}(\sigma_\psi(y))) \; ,
\]
where the mean and log-variance of each distribution are functions parameterized by neural networks. Selecting the parameterized form of these distributions facilitates the reparameterization trick \cite{kingma2014autoencoding}, allowing backpropagation through sampling operations during training. This selection of the prior, encoding, and decoding distributions allows a closed form solution to compute $\mathcal{L}_{VAE}$.

\subsection{Disentanglement}\label{sec:disentanglement}
Disentanglement is realized when variations in a single latent dimension correspond to variations in a single generative parameter. This allows the latent space to be interpretable by the user and improves transferability of representations between tasks. Disentanglement may not be required for some tasks which may not require knowledge on each parameter individually or perhaps only a subset of the generative parameters. Nevertheless, a disentangled representation can be leveraged across many tasks, making it the most comprehensive approach. 

Many metrics of disentanglement exist in the literature \cite{locatello2019challenging}, few of which take into account the generative parameter data. Often knowledge on the generative parameters is lacking, and these metrics can be used to evaluate disentanglement in that case (although there is no consensus on which metric is appropriate). In controlled experiments, however, knowledge on generative parameters is available, and correlation between the latent space and the generative parameter space can be directly determined. To evaluate disentanglement in a computationally efficient manner, we propose a  disentanglement score
\begin{equation} \label{eq:disentanglement_score}
    S_D = \frac{1}{n} \sum_i \frac{\max_j \;\; \textrm{cov}(z_i, \theta_j)}{\sum_j \textrm{cov}(z_i, \theta_j)} \; , 
\end{equation}
where $S_D \in [0.5, 1]$ and scores closer to 1 indicate better disentanglement. It is noted that this score is not used {\em during} the training process. This score is created from the intuition that each latent parameter should be correlated to only a single generative parameter. One might note some issues with this disentanglement score. For instance, if multiple latent dimensions are correlated to the same generative parameter dimension, the score will be inaccurate. Similarly, if the latent dimension is greater than the generative parameter dimension, some latent dimensions may contain no information about the data and be uncorrelated to all dimensions, inaccurately reducing the score. For the cases presented here (we will use the score only when $n = p$), Eq (\ref{eq:disentanglement_score}) suffices as a reasonable measure of disentanglement. This score is used as an efficient means of scoring disentanglement when efficiency is important, but we propose another score based on comparisons between disentangled and entangled representations.

We observed empirically that disentanglement is highly correlated to a match in shape between the generative parameter distribution $p(\theta)$ and the aggregated posterior $q_\phi(z)$ (Section \ref{sec:KLE2}). A match in the scaled-and-translated shapes results in good disentanglement but an aggregated posterior which does not match the shape of the generative parameter distribution or contains incorrect correlations ('rotated') relative to the generative parameter distribution does not. Using this knowledge, another disentanglement metric is postulated to compare these shapes by leveraging the KL Divergence (Eq. \ref{eq:KL_dis}) where $\circ$ denotes the Hadamard product.  
\begin{equation}\label{eq:KL_dis}
    S_{KL} = \min_{a, b} D_{KL}[p(\theta)||q_\phi(a\circ (z - b))].
\end{equation}
This metric compares the shapes of the two distributions by finding the minimum KL divergence between the generative parameter distribution and a scaled and translated version of the aggregated posterior. When $q_\phi(a \circ (z-b))$ is close to $p(\theta)$ for some vectors $a, b \in \mathbb{R}^n$, disentanglement is observed.

It is noted in \cite{rolinek2019variational} that rotation of the latent space certainly has a large effect on disentanglement, which is precisely what we observe (Section \ref{sec:KLE2}). Additionally, the ELBO loss is unaffected by rotations of the latent space when using rotationally-invariant priors such as standard normal (\ref{app:rot_inv}).  

\subsubsection{$\beta$-VAE}\label{sec:beta_VAE}
The $\beta$-VAE objective gives greater weighting to the regularization loss, 
\begin{equation}\label{eq:beta_VAE_loss}
    \mathcal{L}_{\beta-VAE} = \beta\mathbb{E}_{\hat{p}(y)}[D_{KL}[q_\phi(z|y)||p(z)] +  \mathbb{E}_{\hat{p}(y)}[\mathbb{E}_{q_\phi(z|y)}[-\log p_\psi(y|z)]] \;.
\end{equation}
This encourages greater regularization, often leading to improved disentanglement over the standard VAE loss \cite{Higgins2017betaVAELB}. It is worth noting that when $\beta$ = 1, with a perfect encoder and decoder, the VAE loss reduces to the Bayes rule \cite{yu2020tutorial,duraisamy2021variational}. More details on the $\beta$-VAE are provided in Section \ref{sec:RD}.

\subsection{Connections to Rate-Distortion Theory} \label{sec:RD}
Rate-distortion theory \cite{RDberger,RDcover,RDgibson} aids in a deeper understanding in the trade off and balance between the regularization and reconstruction losses. The general rate distortion problem is formulated before making these connections.

Consider two random variables: data $Y: \Omega \rightarrow \mathbb{R}^n$ and a compressed representation of the data $Z: \Omega \rightarrow \mathbb{R}^p$. An encoder $p(z|y)$ is sought such that the compressed representation contains a minimal amount of information about the data subject to a bounded error in reconstructing the data.

A model $\tilde{y}(z)$ is used to reconstruct $Y$ from samples of $Z$, and a distortion metric $d(y, \tilde{y})$ is used as a measure of error in the reconstruction of $Y$ with respect to the original data. 

A rate-distortion problem thus takes the general form
\begin{equation}\label{eq:RD}
    R(D) = \min_{p(z|y)} I(Y;Z) \;\;\; \textrm{s.t.}\;\; \mathbb{E}_{Y,Z}[d(y, \tilde{y}(z))] \leq D \; ,
\end{equation}
where $D \in \mathbb{R}$ is an upper bound on the distortion. Solutions to Eq \ref{eq:RD} consist of an encoder $p(z|y)$ which extracts as little information as possible from $Y$ while maintaining a bounded distortion on the reconstruction of $Y$ from $Z$ through the model $\tilde{y}(z)$. Mutual information is minimized to obtain a maximally compressed representation of the data. Learning unnecessary information leads to 'memorization' of some aspects of the data rather than extracting only the information relevant to the task at hand. 

This optimization problem formulated as the rate-distortion Lagrangian is
\begin{equation}\label{eq:RD_lagrangian}
    \min \mathcal{J}(\beta) = \min_{p(z|y)} I(Y;Z) + \beta(\mathbb{E}_{Y,Z}[d(y, \tilde{y}(z)]-D) \; .
\end{equation}

\begin{figure}[h!]
    \centering
    \captionsetup{width=.6\linewidth}
    \includegraphics[width=.6\textwidth,angle=0,clip,trim=0pt 0pt 0pt 0pt]{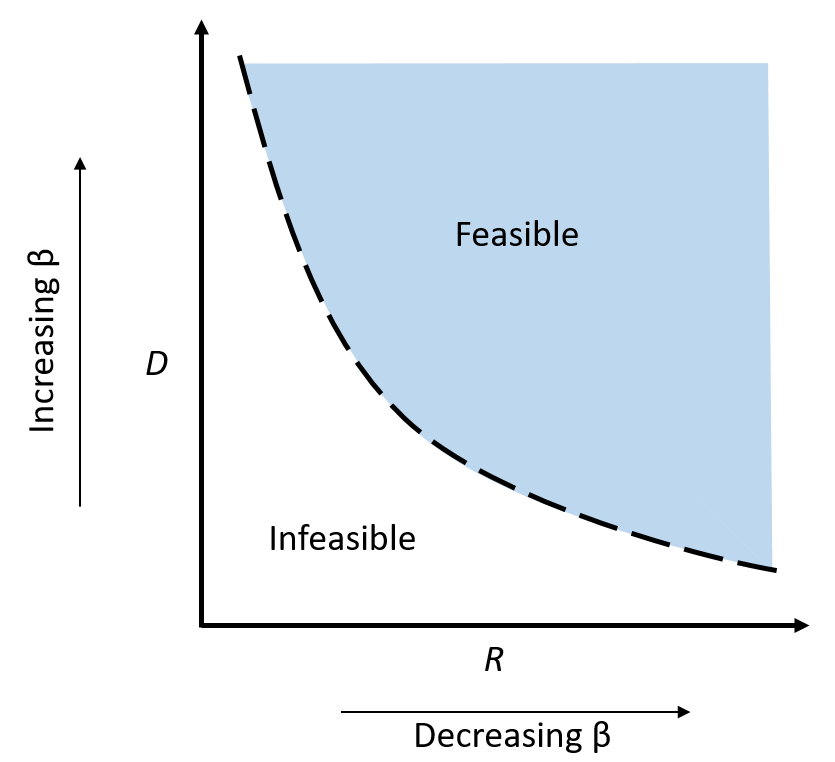}
    \caption{Convex rate distortion (RD) curve.}
    \label{fig:RD_curve}
\end{figure}

Given an encoder and decoder, solutions to the rate-distortion problem lie on a convex curve referred to as the rate-distortion curve \cite{Alemi2018FixingAB}. Points above this curve correspond to \emph{realizable} yet sub-optimal solutions. Points below the RD curve correspond to solutions which are \emph{not} realizable; no possible compression exists with distortion below the RD curve. As the RD curve is convex, optimal solutions found by varying $\beta$ lie along the curve. Increasing $\beta$ increases the tolerable distortion, decreasing the mutual information between the compressed representation and data, providing a more compressed representation. Conversely, decreasing $\beta$ requires a more accurate reconstruction of the data, increasing the mutual information between compressed representation and data. Figure \ref{fig:RD_curve}  illustrates the concept of the RD curve. 

The $\beta$-VAE loss is tied to a rate-distortion problem. Rearranging the VAE regularization loss ($L_{REG}$), we obtain
\[
L_{REG} = \mathbb{E}_{\hat{p}(y)}[D_{KL}[q_\phi(z|y)||p(z)] = I_\phi(Y;Z) + D_{KL}[q_\phi(z)||p(z)] \; .
\]
Minimizing the $\beta$-VAE loss gives an optimization problem
\[
\min_{\phi, \psi} L_{\beta-VAE} = \min_{\phi, \psi} I_\phi(Y;Z) + D_{KL}[q_\phi(z)||p(z)] + \beta \mathbb{E}_{\hat{p}(y)p_\psi(z|y)}[-\log p_\phi(y|z)] \; .
\]
This optimization problem is similar to a rate-distortion problem with $d(y, \tilde{y}) = -\log p_\phi(y|z)$, but contains an additional KL-Divergence term between the aggregated posterior and the prior. Additionally, $I_\phi(Y;Z)$ is an approximation to the true mutual information $I(Y;Z)$. Depending on $\beta$, solutions can be found at any location along the RD curve with each containing differing properties. RD curve for VAEs is simply an analogy: $L_{REG}$ is considered the rate $R$ and $L_{REC}$ is considered the distortion $D$. 

With increased $\beta$, the $\beta$-VAE encourages $q_\phi(z)$ to be closer to $p(z)$ but also lowers the mutual information between the data and the latent parameters, limiting reconstruction accuracy. In \cite{rolinek2019variational}, disentanglement is illustrated to be caused inadvertently through the assumed factored form of the encoding distribution even though rotations of the latent space have no effect on the ELBO. However, their proof relies on training in the 'polarized' regime characterized by loss of information or 'posterior collapse' \cite{Lucas2019UnderstandingPC}. Training in this regime often requires increasing the weight of the regularization loss, necessarily decreasing reconstruction performance in the process. In our work, we illustrate disentanglement through training VAEs with the ELBO loss ($\beta$ = 1), keeping reconstruction accuracy high. Ref.~\cite{rolinek2019variational} presents good insight in disentanglement. 

\subsection{Hierarchical Priors} \label{sec:hierarchical_priors}
Often the prior (in the case of classic VAEs, specified by the user) and generative parameter distributions (data dependent) may not be highly correlated. Hierarchical priors \cite{klushyn2019learning} (HP) can be implemented within the VAE network such that the prior is learned as a function of additional random variables, potentially leading to more expressive priors and aggregated posteriors. Hierarchical random variables $\xi_i$ are introduced such that 'sub-priors' can be assumed on each $\xi_i$ (typically standard normal). In the case of a single hierarchical random variable 
\[
p(z) = \int_{\Xi} p(z|\xi)p(\xi) d\xi = \int_{\Xi} \frac{p(\xi|z)}{p(\xi|z)}p(z|\xi)p(\xi)d\xi = \mathbb{E}_{\Xi \sim p(\xi|z)} \left[ \frac{p(z|\xi)p(\xi)}{p(\xi|z)} \right ]\; .
\]
The conditional distributions $p(\xi|z)$ and $p(z|\xi)$ are the \emph{prior encoder} and \emph{prior decoder}, respectively. These distributions can be approximated by parameterizing them with neural networks. The parameterized distributions are noted as $q_\gamma(\xi|z)$ and $p_\pi(z|\xi)$ where $\gamma$ are the trainable parameters of the approximating prior encoder and $\pi$ are the trainable parameters of the prior decoder. Thus, the VAE prior can be approximated through the prior encoding and decoding distributions
\begin{equation} \label{eq:hierarchical_prior}
    p(z) \approx \mathbb{E}_{\Xi \sim q_\gamma(\xi|z)} \left[ \frac{p_\pi(z|\xi)p(\xi)}{q_\gamma(\xi|z)} \right ] \; .
\end{equation}
Rearranging the VAE regularization loss 
\begin{equation}\label{eq:reg_loss_expansion}
    L_{REG} = \int_{Y,Z} p(y)q_\phi(z|y) \log \frac{q_\phi(z|y)}{p(z)} dydz = \mathbb{E}_{Y,Z\sim p(y)q_\phi(z|y)}[\log q_\phi(z|y)] - \int_{Y,Z} p(y)q_\phi(z|y)\log p(z) dydz \; ,
\end{equation}
and substituting the approximating hierarchical prior (\ref{eq:hierarchical_prior}) into (\ref{eq:reg_loss_expansion}), the final term on the right-hand side becomes
\[
 - \int_{Y,Z} p(y)q_\phi(z|y)\log p(z) dydz = - \int_{Y,Z} p(y)q_\phi(z|y) \log \left [ \mathbb{E}_{\Xi \sim q_\gamma(\xi|z)} \left[ \frac{p_\pi(z|\xi)p(\xi)}{q_\gamma(\xi|z)} \right ] \right ] dydz \; .
\]
The logarithm function is strictly concave; therefore, by Jensen's inequality the right-hand side is upper bounded by 
\[
- \int_{Y,Z} p(y)q_\phi(z|y) \log \left [ \mathbb{E}_{\Xi \sim q_\gamma(\xi|z)} \left[ \frac{p_\pi(z|\xi)p(\xi)}{q_\gamma(\xi|z)} \right ] \right ] dydz \leq - \int_{Y,Z} p(y)q_\phi(z|y) \mathbb{E}_{\Xi \sim q_\gamma(\xi|z)} \left [ \log \frac{p_\pi(z|\xi)p(\xi)}{q_\gamma(\xi|z)} \right ] dydz \; .
\]
This bound is rearranged to the form 
\begin{equation} \label{eq:prior_vae_loss}
    \mathbb{E}_{Y,Z\sim p(y)q_\phi(z|y)}[D_{KL}[q_\gamma(\xi|z)||p(\xi)] - \mathbb{E}_{Y,Z\sim p(y)q_\phi(z|y)}[\mathbb{E}_{q_\gamma(\xi|z)}[\log p_\pi(z|\xi)]] \;.
\end{equation}
Equation \ref{eq:prior_vae_loss} takes the same form as the overall VAE loss, but applied to the prior network itself. Thus, the hierarchical prior can be thought of as a system of sub-VAEs within the main VAE. In summary, the VAE loss is upper bounded by
\begin{align}\label{eq:vae_hp_loss}
    L_{VAE} & \leq  \mathbb{E}_{Y,Z\sim \hat{p}(y)q_\phi(z|y)}[\log q_\phi(z|y)] + \mathbb{E}_{Y,Z\sim \hat{p}(y)q_\phi(z|y)}[D_{KL}[q_\gamma(\xi|z)||p(\xi)] \\
    & - \mathbb{E}_{Y,Z\sim \hat{p}(y)q_\phi(z|y)}[\mathbb{E}_{q_\gamma(\xi|z)}[\log p_\pi(z|\xi)]] - \mathbb{E}_{Y,Z \sim \hat{p}(y)q_\phi(z|y)}[\log p_\psi(y|z)]] \;. 
\end{align}
Note that the expectation with respect to $p(y)$ is replaced by an empirical expectation over the data.

Implementing hierarchical priors can aid in learning non-rotationally-invariant priors, frequently inducing a learned disentangled representation, as shown below.

\section{Sample Application: Darcy Flow} \label{sec:darcy_flow}
To characterize the training process of the VAEs and to study disentanglement, we employ an application of flow through  porous media. A two-dimensional steady-state Darcy flow problem in $s$ spatial dimensions (our experiments employ $s=2$) is governed by \cite{Zhu_2018} 
\begin{align}\label{darcy}
    u(x) & = -K(x)\nabla p(x), \;\;\; x \in \mathcal{X} \nonumber \\
    \nabla \cdot u(x) & = f(x), \;\;\; x \in \mathcal{X} \\
    u(x) \cdot \hat{n}(x) & = 0, \;\;\; x \in \partial\mathcal{X} \nonumber \\
    \int_{\mathcal{X}} p(x)dx & = 0. \nonumber
\end{align}

Darcy's law is an empirical law describing flow through porous media in which the permeability field is a function of the spatial coordinate $K(x) : \mathbb{R}^{s} \rightarrow \mathbb{R}$. The pressure $p(x) : \mathbb{R}^s \rightarrow \mathbb{R}$ and velocity $u(x) : \mathbb{R}^s \rightarrow \mathbb{R}^s$ are found given the source term $f(x) : \mathbb{R}^s \rightarrow \mathbb{R}$, permeability, and boundary conditions. The integral constraint is given to ensure a unique solution. 

A no-flux boundary condition is specified, and the source term models an injection well in one corner of the domain and a production well in the other 
\begin{equation}\label{darcy_source}
f(x) = 
\begin{cases}
    r, & \;\; |x_i - \frac{1}{2}w|\leq \frac{1}{2}w, \;\; i = 1,2 \\
    -r, & \;\; |x_i-1+\frac{1}{2}w|\leq \frac{1}{2}w, \;\; i = 1,2 \\
    0, & \;\; \textrm{otherwise}
\end{cases} \; \; \; , 
\end{equation}
where $w = \frac{1}{8}$ and $r = 10$. The computational domain considered is the unit square $\mathcal{X} = [0,1]^2$. 

\subsection{KLE Dataset}
The dataset investigated uses a log-permeability field modeled by a Gaussian random field with covariance function $k$
\begin{equation} \label{permeability}
    K(x) = \mathrm{exp}(G(x)), \;\; G(\cdot) \sim \mathcal{N}(m, k(\cdot,\cdot)) \; . 
\end{equation}
Generating the data first requires sampling from the permeability field (Eq. (\ref{permeability})). We use Eq \ref{eq:cov_func} as the covariance function in our experiments, as in \cite{Zhu_2018}. 
\begin{equation} \label{eq:cov_func}
    k(x,x') = \textrm{exp}(-||x-x'||_2/l)
\end{equation}
After sampling the permeability field, solving Eq. (\ref{darcy}) for the pressure and velocity fields produces data samples. We discretize the spatial domain on a $65\times 65$ grid and use a second-order finite difference scheme to solve the system.

The intrinsic dimensionality of the data will be the total number of nodes in the system ($4225$ for our system) \cite{Zhu_2018}. For dimensionality reduction, the intrinsic dimensionality $p$ of the data is specified by leveraging the Karhunen-Loeve Expansion (KLE), keeping only the first $p$ terms in
\begin{equation} \label{KLE}
    G(x) = m + \sum_{i=1}^p \sqrt{\lambda_i} \theta_i \phi_i(x) \; ,
\end{equation}
where $\lambda_i$ and $\phi_i(x)$ are eigenvalues and eigenfunctions of the covariance function (Eq. \ref{eq:cov_func}) sorted by decreasing $\lambda_i$, and each $\theta_i$ are sampled according to some distribution $p(\theta)$, denoted the \emph{generative parameter distribution}.

\begin{figure}[h!]
    \centering
    \captionsetup{width=.8\linewidth}
    \includegraphics[width=.49\textwidth,angle=0,clip,trim=0pt 0pt 0pt 0pt]{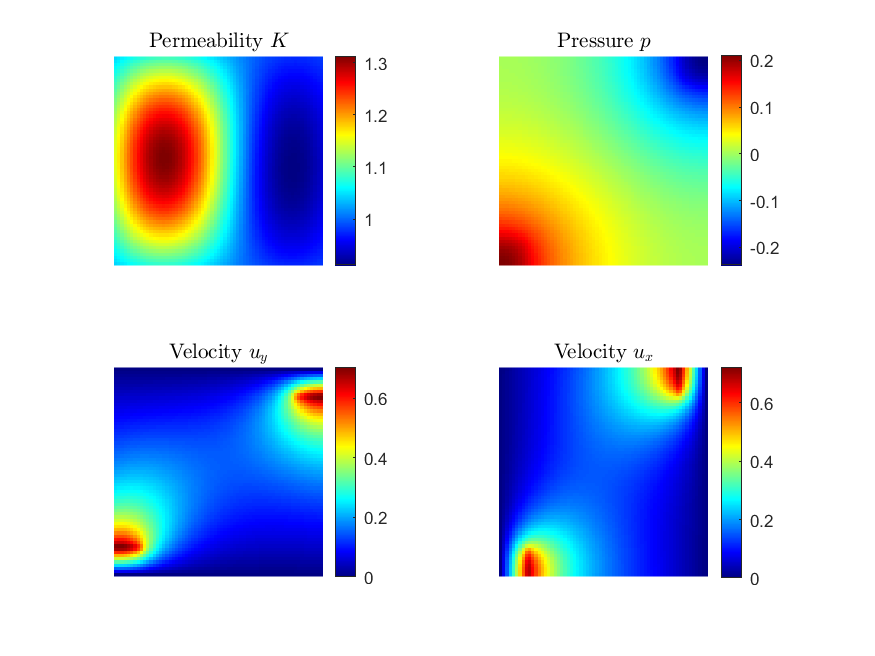}
    \includegraphics[width=.49\textwidth,angle=0,clip,trim=0pt 0pt 0pt 0pt]{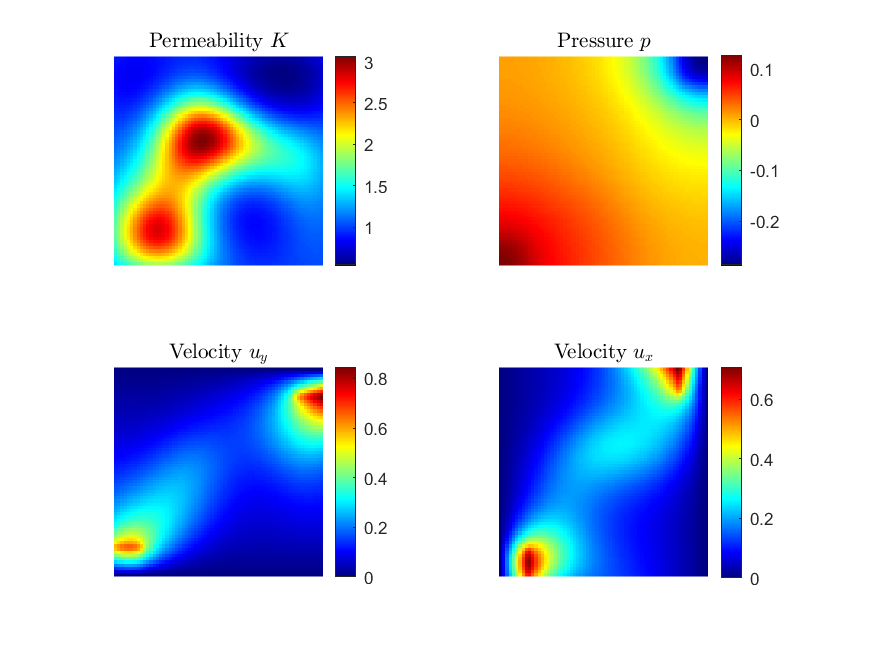}
    \includegraphics[width=.49\textwidth,angle=0,clip,trim=0pt 0pt 0pt 0pt]{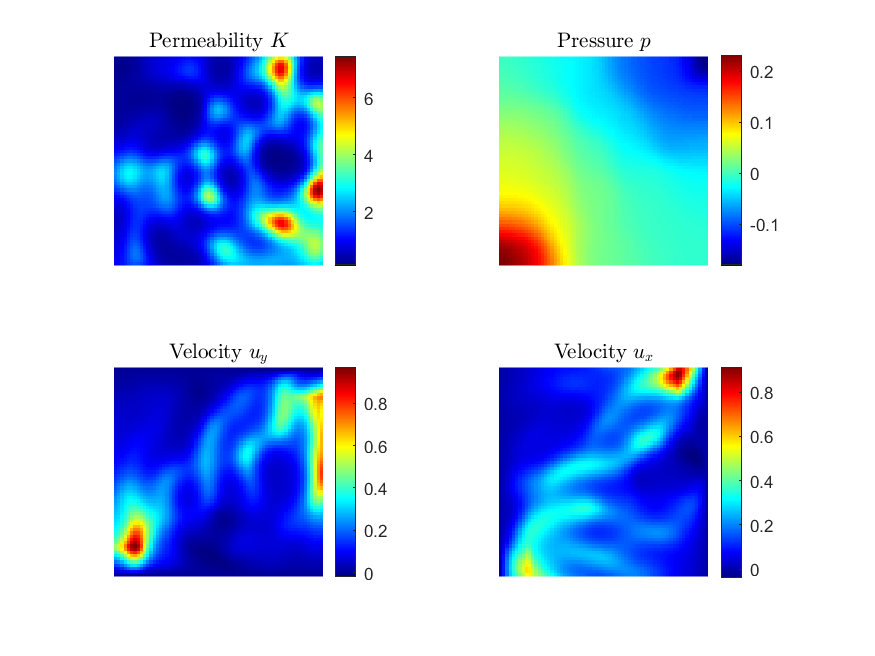}
    \includegraphics[width=.49\textwidth,angle=0,clip,trim=0pt 0pt 0pt 0pt]{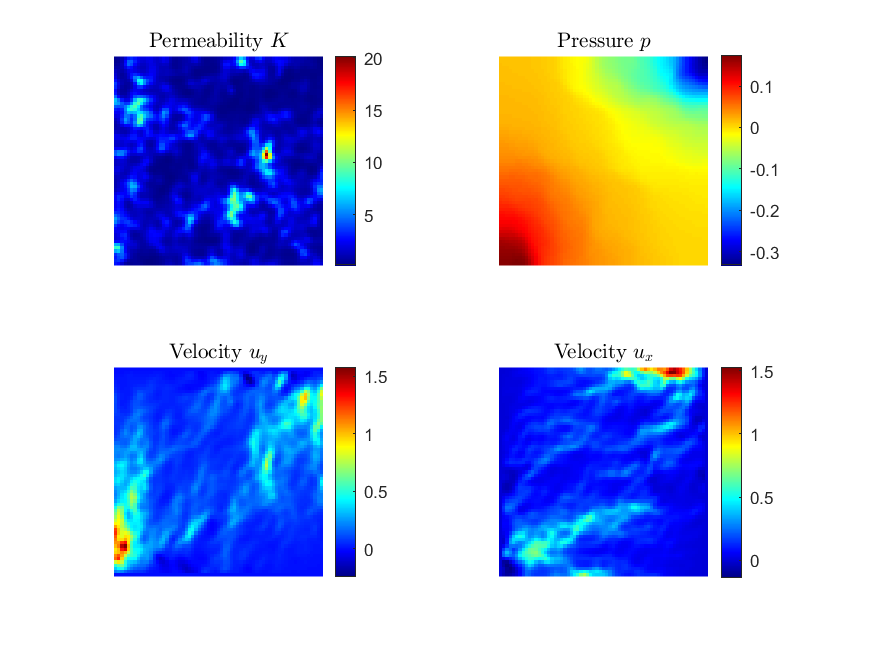}
    \caption{Samples from datasets (\emph{top left}) KLE2 (\emph{top right}) KLE10 (\emph{bottom left}) KLE100 (\emph{bottom right}) KLE1000.}
    \label{fig:datasets}
\end{figure}

Each dataset contains some intrinsic dimensionality $p$, and we denote each dataset using the permeability field (Eq \ref{permeability}) as KLE$p$. For example, a dataset with $p=100$ is referred to as KLE100. Samples from datasets of various intrinsic dimension are illustrated in Figure \ref{fig:datasets}. 
Variations on the KLE2 dataset are employed for our explorations in this work. The differences explored are related to varying the generative parameter distribution $p(\theta)$ in each set. 

Each snapshot $y_i$ from a single dataset $Y = \{y_i\}_{i=1}^N$ contains the pressure $p(x)$ and velocity fields $u(x)$ at each node in the computational domain. These are used as a 3-channel input to the VAE; the permeability field and KLE expansion coefficients (generative parameters) are saved and used only for evaluation purposes.

\section{Training Setup and Loss Landscape}
\label{sec:issues}

The process of training a VAE involves a number of challenges. For example, convergence of the optimizer to local minima can greatly hinder reconstruction accuracy and failure to converge altogether remains a possibility. A recurrent issue with VAE training in our experiments is that of over-regularization. Over-regularized solutions are characterized by disproportionately small regularization loss ($L_{REG} << 1$). More information on this issue is detailed in Section \ref{sec:over_reg}.  

To mitigate some of the issues inherent to training VAEs, we employ a training method tailored to avoid over-regularization. The model is trained initially with $\beta_0 << 1$, typically around $\beta_0 = 10^{-7}$, for some number of epochs $r_0$ (depending on learning rates) until reconstruction accuracy is well below that of an over-regularized solution (Section \ref{fig:over_reg} illustrates this necessity). When $\beta_0$ is too small, the regularization loss can become too large, preventing convergence altogether. Training is continued by implementing a $\beta$ scheduler~\cite{klushyn2019learning} to slowly increase the weight of the regularization loss.  The learning rate is then decreased to $lr_1 = c (lr_0)$ after some number of epochs $r_1$ to enhance reconstruction accuracy. This training method - in particular the heavily weighted reconstruction phase and the $\beta$ scheduler - result in much more stable training which avoids the local minima characterized by over-regularization and improves convergence consistency. Similar methods have been employed to avoid this issue. In particular, \cite{fu2019cyclical} refers to this issue as ``KL vanishing" and uses a cyclical $\beta$ schedule to avoid the issue. However, this can take far more training epochs and cycle iterations to converge than the method employed here. 

\subsection{Architecture} \label{sec:architecture}

\begin{figure}[h!]
    \centering
    \captionsetup{width=.8\linewidth}
    \includegraphics[width=.95\textwidth,angle=0,clip,trim=0pt 0pt 0pt 0pt]{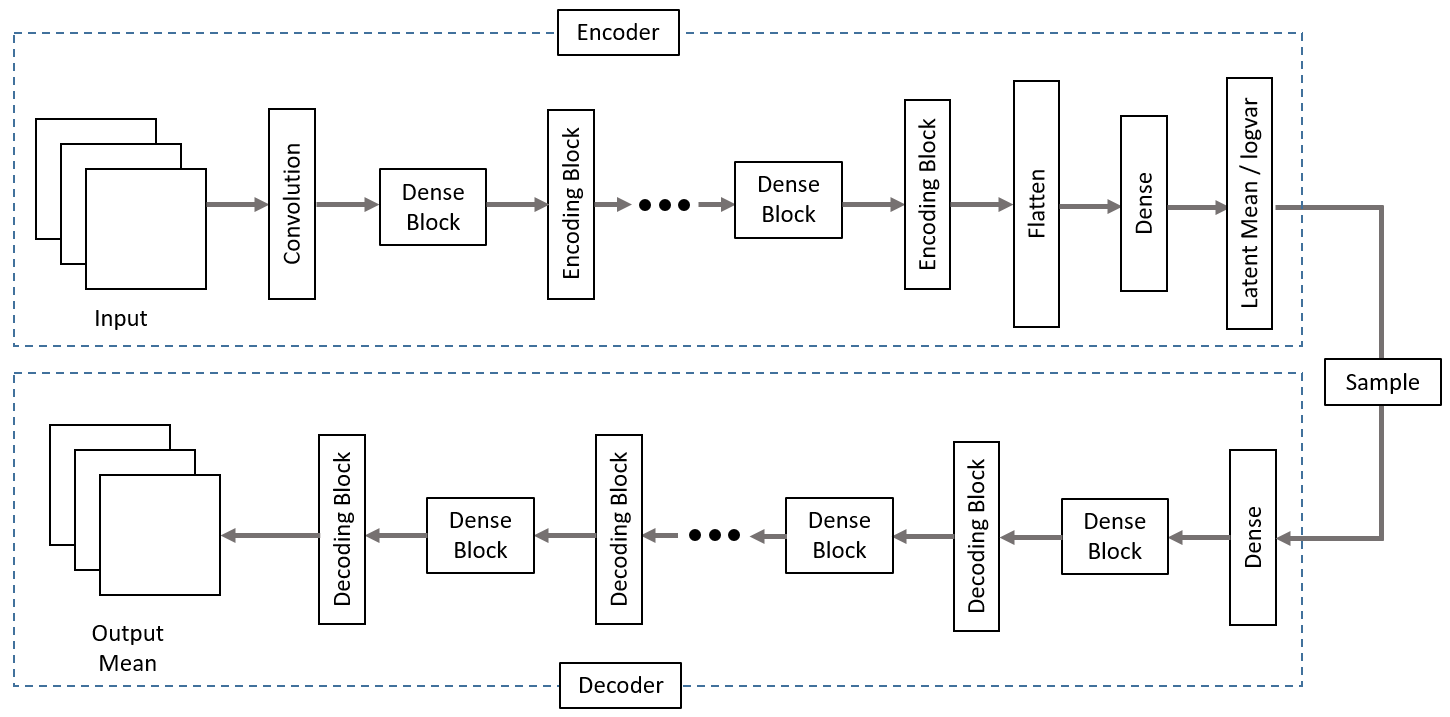}
    \caption{Dense VAE architecture.}
    \label{fig:dense_architecture}
\end{figure}

The architecture for the VAE is adapted from Ref.~\cite{Zhu_2018}. A convolutional layer is first applied to the input. A series of dense blocks and encoding blocks followed by a flatten and fully connected layers then encode the input to the parameterized latent distribution. A series of decoding blocks and dense blocks then decode samples from the latent space to the output. Figure \ref{fig:dense_architecture} illustrates the general architecture we have selected for the latent mean and log-variance along with the output mean (with the output log-variance being constant but trainable). 

\begin{figure}[h!]
    \centering
    \captionsetup{width=.8\linewidth}
    \includegraphics[width=.49\textwidth,angle=0,clip,trim=0pt 0pt 0pt 0pt]{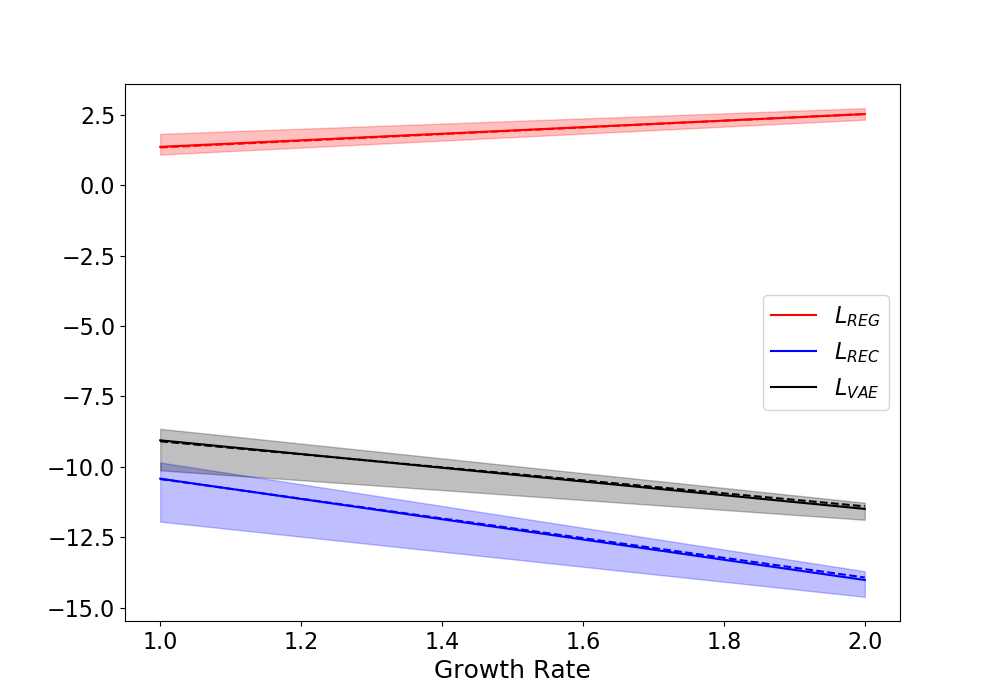}
    \includegraphics[width=.49\textwidth,angle=0,clip,trim=0pt 0pt 0pt 0pt]{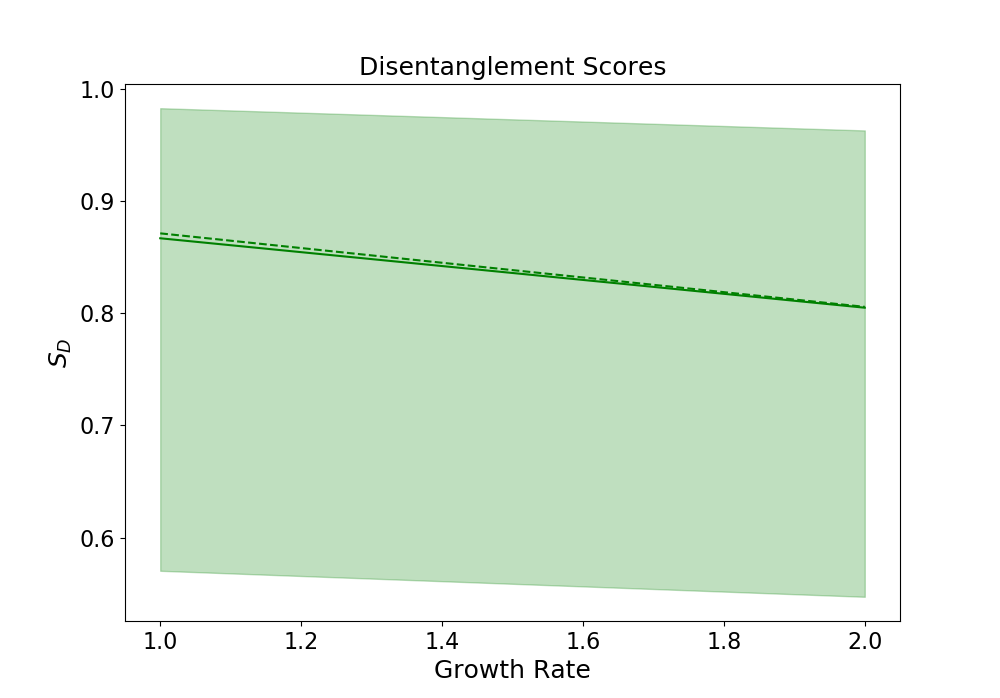}
    \includegraphics[width=.49\textwidth,angle=0,clip,trim=0pt 0pt 0pt 0pt]{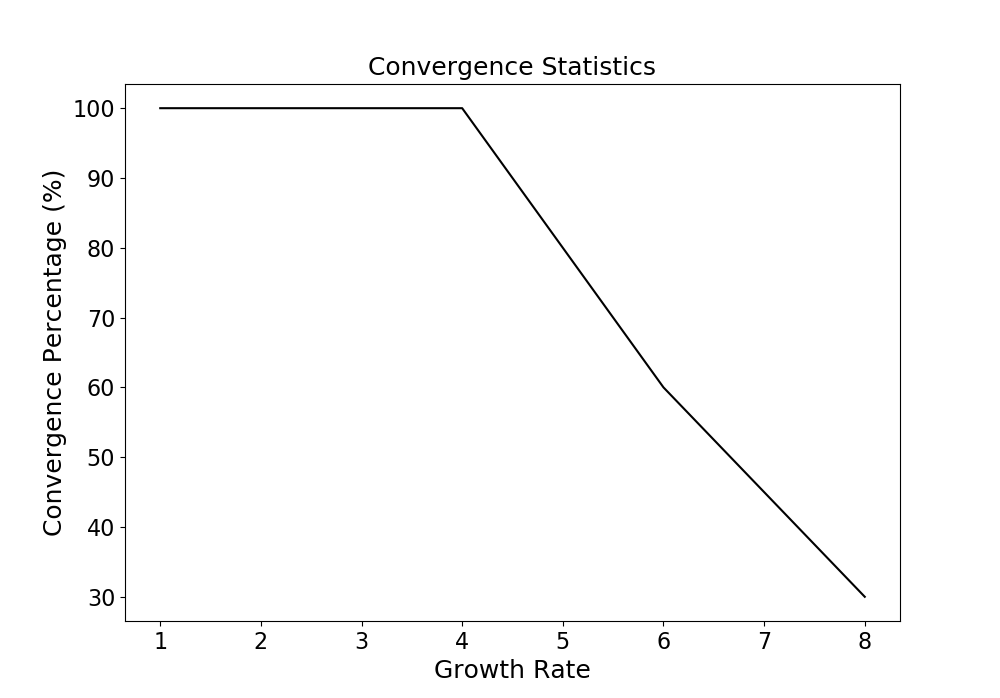}
    \caption{Hyperparameter selection example.}
    \label{fig:growth_rate_selection}
\end{figure}

A convolutional architecture was initially implemented without the use of dense blocks, but reconstruction of data (Section \ref{sec:darcy_flow}) was not very accurate with this architecture, even with some amount of hyperparameter tuning. Ref.~\cite{Zhu_2018} illustrated that the architecture implemented there can accurately predict the data of our problem. The architecture contains many hyperparameters such as number of dense blocks, number of layers in each dense block, dense block growth rate, stride of convolutions, fully connected layer width, and others. There were three main goals for us in tuning the hyperparameters: accurate reconstruction, ability to produce disentangled representations, and high computational efficiency. As an example of hyperparameter tuning, we consider changes in the dense block growth rate keeping all other hyperparameters constant. Ten VAEs were trained with the ELBO loss for each growth rate value on the KLE2 dataset with $p(\theta)$ being standard normal. Figure \ref{fig:growth_rate_selection} illustrates some statistics on this study. The overall ELBO loss, and in particular the reconstruction loss continues to decrease with and increase in growth rate, which is desirable. Good conclusions cannot be easily drawn from the disentanglement statistics, although at each growth rate a disentangled representation was observed. However, as the growth rate increases, the probability of convergence decreases. This may be improved by introducing lower learning rates, but in our case increase training time was highly undesirable. Thus, a growth rate of 4 was selected.  

This architecture is used for all VAEs trained in this work. The latent and output distributions are assumed to be Gaussian. We use the dense architecture to parameterize the encoder mean and log-variance separately, as well as the decoder mean. The decoding distribution log-variance is learned but constant as introducing a learned output log-variance did not aid in reconstruction or improving disentanglement properties in our experiments but increased training time. 

\subsection{Over-Regularization} \label{sec:over_reg}
\begin{figure}[h!]
    \centering
    \captionsetup{width=.8\linewidth}
    \includegraphics[width=.99\textwidth,angle=0,clip,trim=0pt 0pt 0pt 0pt]{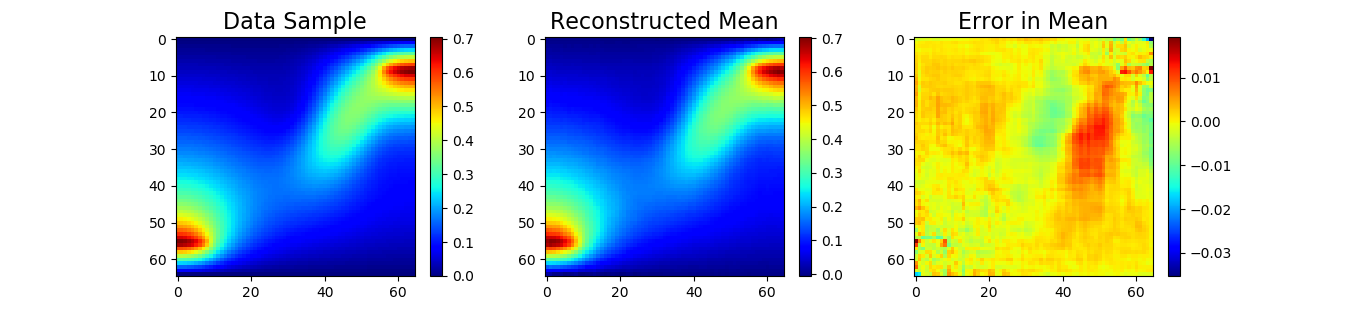}
    \includegraphics[width=.97\textwidth,angle=0,clip,trim=0pt 0pt 0pt 0pt]{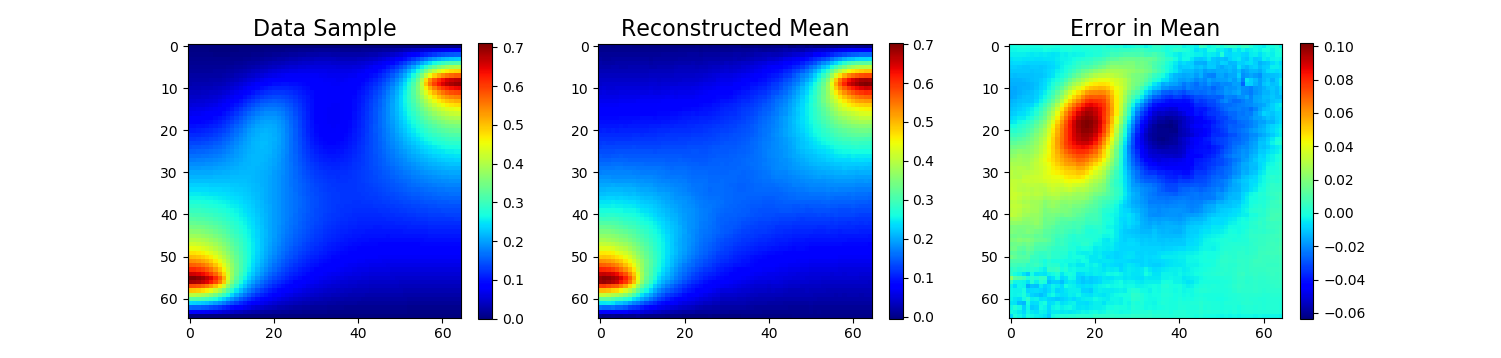}
    \caption{(\emph{upper}) Good reconstruction. (\emph{lower}) Over-regularization.}
    \label{fig:recon}
\end{figure}

Over-regularization has been identified as a challenge in the  training of VAEs (e.g. \cite{fu2019cyclical}). This phenomenon is characterized by the latent space containing no information about the data; i.e. the regularization loss becomes zero. The decoder of the VAE learns to predict a constant output (it does not depend on the latent variable). The constant output it learns becomes a normal distribution with mean and variance of the data. With zero regularization loss, the learned decoding distribution becomes $p_\psi(y|z) =  \mathcal{N}(y; \hat{\mu}_y, \textrm{diag}(\hat{\sigma}^2_y))$ $\forall$ $z$ where $\hat{\mu}_y = \frac{1}{N}\sum_{i=1}^N y^{(i)}$ and $\hat{\sigma}^2_y = \frac{1}{N}\sum_{i=1}^N(y^{(i)}-\hat{\mu}_y)^2$. This is proven to minimize $L_{VAE}$ in Theorem \ref{thm:over_reg}. The solution is not shown to be unique, but our experiments indicate that this is the over-regularized solution found during training. As Theorem \ref{thm:over_reg} illustrates validity for any encoder $q(z|y)$, this is the most robust solution for the VAE to converge to when over-regularization occurs. The decoder learns to predict as accurately as possible given nearly zero mutual information between the latent and data random variables. As the encoder and decoder are trained simultaneously, predicting a constant output regardless of $z$ prevents the necessity of the decoder to adjust as the encoder changes. An empirical comparison between good reconstruction and over-regularization is shown in Figure \ref{fig:recon}. 

Theorem \ref{thm:over_reg} requires that the output variance is constant. Parameterizing the output variance with an additional network may aid in avoiding over-regularization. 

\begin{theorem} \label{thm:over_reg} 
Given data $\{ y^{(i)} \}_{i=1}^N$ and the VAE framework defined in Section \ref{sec:vae}, 
and assuming a decoding distribution of the form $p(y|z) = \mathcal{N}(y; \mu(z), \textrm{diag}(\sigma^2))$, 
if $L_{REG} = 0$, then $\textrm{argmin}_{\mu(z), \sigma^2}\;\; L_{VAE} = \{\hat{\mu}_y, \hat{\sigma}^2_y\}$ , where $\hat{\mu}_y = \frac{1}{N}\sum_{i=1}^N y^{(i)}$ and $\hat{\sigma}^2_y = \frac{1}{N}\sum_{i=1}^N(y^{(i)}-\hat{\mu}_y)^2$.

\end{theorem}

\begin{proof}
For any $q(z|y)$ s.t. $L_{REG} = 0$:
\begin{align}
    L_{VAE} = L_{REC} = & \mathbb{E}_{\hat{p}(y)q(z|y)}[-\log(p(y|z))] \\
    = &\mathbb{E}_{q(z|y)}\left [ \frac{1}{N} \sum_{i=1}^N \sum_{j=1}^m \frac{1}{2}\log(2\pi) + \log(\sigma_j) + \frac{1}{2\sigma^2_j}(y_j^{(i)} - \mu_j(z))^2 \right ].
\end{align}

To minimize $L_{VAE}$, take derivatives $\frac{\partial L_{VAE}}{\partial \mu_j(z)}$ and $\frac{\partial L_{VAE}}{\partial \sigma_j}$ (assuming derivative and expectation can be interchanged):
\[
\frac{\partial L_{VAE}}{\partial \mu_j(z)} = \mathbb{E}_{q(z|y)} \left [ -\frac{1}{N}\sum_{i=1}^N \frac{1}{\sigma_j^2}(y^{(i)}_j-\mu_j(z)) \right ] = 0. 
\]
Thus,
\[
\mathbb{E}_{q(z|y)} \left [\sum_{i=1}^N y^{(i)}_j-\mu_j(z)\right ] = 0
\]
and 
\begin{equation}\label{eq:mu_statement}
    \mathbb{E}_{q(z|y)}[\mu_j(z)] = \frac{1}{N}\sum_{i=1}^N y_j^{(i)}.
\end{equation}
Setting
\begin{equation} \label{eq:over_reg_mu}
    \mu_j(z) = \hat{\mu}_j = \frac{1}{N} \sum_{i=1}^N y^{(i)}_j
\end{equation}
Eq \ref{eq:mu_statement}  holds $\forall \; z, \; j$. 

Taking the derivative w.r.t. variance:
\[
\frac{\partial L_{VAE}}{\partial \sigma_j} = \mathbb{E}_{q(z|y)} \left [ \frac{1}{N}\sum_{i=1}^N \frac{1}{\sigma_j} -  \frac{1}{\sigma_j^3}(y^{(i)}-\mu_j(z))^2 \right ] = 0
\]
Rearranging:
\begin{equation}\label{eq:over_reg_sigma}
    \sigma_j^2 = \frac{1}{N} \sum_{i=1}^N (y_j^{(i)} - \mu_j(z))^2 \;\; \forall \;\; z, j.
\end{equation}
Substituting Eq \ref{eq:mu_statement} into Eq \ref{eq:over_reg_sigma} results in:
\begin{equation}\label{eq:sigma_statement}
    \hat{\sigma}_j^2 = \frac{1}{N}\sum_{i=1}^N (y_j^{(i)} - \hat{\mu}_j)^2 \; \forall \; z, \; j
\end{equation}
With \ref{eq:mu_statement} and \ref{eq:sigma_statement} valid for all $z$ and $j$, we can combine them into vector form and note that Eq \ref{eq:min_mu_sigma} minimizes $L_{VAE}$ as required.
\begin{equation}\label{eq:min_mu_sigma}
    \hat{\mu}_y = 
    \begin{bmatrix}
        \hat{\mu}_1 \\
        \vdots \\
        \hat{\mu}_m
    \end{bmatrix} 
    , \ \  \hat{\sigma}^2_y = 
    \begin{bmatrix}
        \hat{\sigma}^2_1 \\
        \vdots \\
        \hat{\sigma}^2_m   
    \end{bmatrix}.
\end{equation}
\end{proof}

There exists a region in the trainable parameter loss landscape characterized by over-regularized local minimum solutions which partially surrounds the 'desirable' solutions characterized by better reconstruction accuracy and latent properties. This local minima region is often avoided by employing the training method discussed previously, but random initialization of network parameters and changes in hyperparameters between training can render it difficult to avoid convergence to this region. 

We illustrate the problem of over-regularization by training VAEs using the architecture described on Section \ref{sec:architecture} on the KLE2 Darcy flow dataset with $p(\theta)$ being standard normal. 

\begin{figure}[h!]
    \centering
    \captionsetup{width=.8\linewidth}
    \includegraphics[width=.49\textwidth,angle=0,clip,trim=0pt 0pt 0pt 0pt]{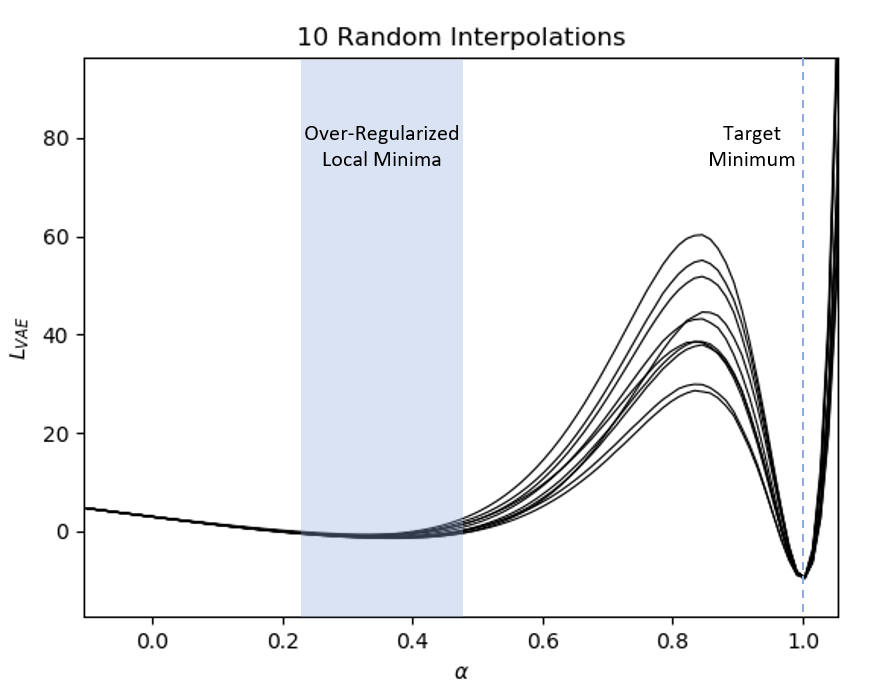}
    \caption{Loss along interpolated lines between 10 random weight initializations and a desirable converged solution.}
    \label{fig:over_reg}
\end{figure}

A VAE is trained with 512 training samples (each sample is 65$\times$65$\times$3), converging to a desirable solution with low reconstruction error and nearly perfect disentanglement. The parameters of this trained network are denoted $P_T$. After the VAE is trained and a 'desirable' solution obtained, 10 additional VAEs with identical setup to the desirable solution are initialized randomly using the Xavier uniform weight initialization on all layers. Each of the 10 initializations contain parameters $P_i$. A line in the parameter space is constructed between the converged 'desirable' solution and the initialized solutions as a function of $\alpha$:

\begin{equation} \label{eq:alpha_line}
    P(\alpha) = (1-\alpha)P_i + \alpha P_T.
\end{equation} 

Losses are recorded along each of the 10 interpolated lines and plotted in Figure \ref{fig:over_reg}. Between the random initializations and 'desirable' converged solutions there exists a region of local minima in the loss landscape, and these local minima are characterized by over-regularization. Losses illustrated are computed as an expectation over all training data and a Monte Carlo estimate of the reconstruction loss with 10 latent samples to limit errors due to randomness. 

\begin{figure}[h!]
    \centering
    \captionsetup{width=.8\linewidth}
    \includegraphics[width=.49\textwidth,angle=0,clip,trim=0pt 0pt 0pt 0pt]{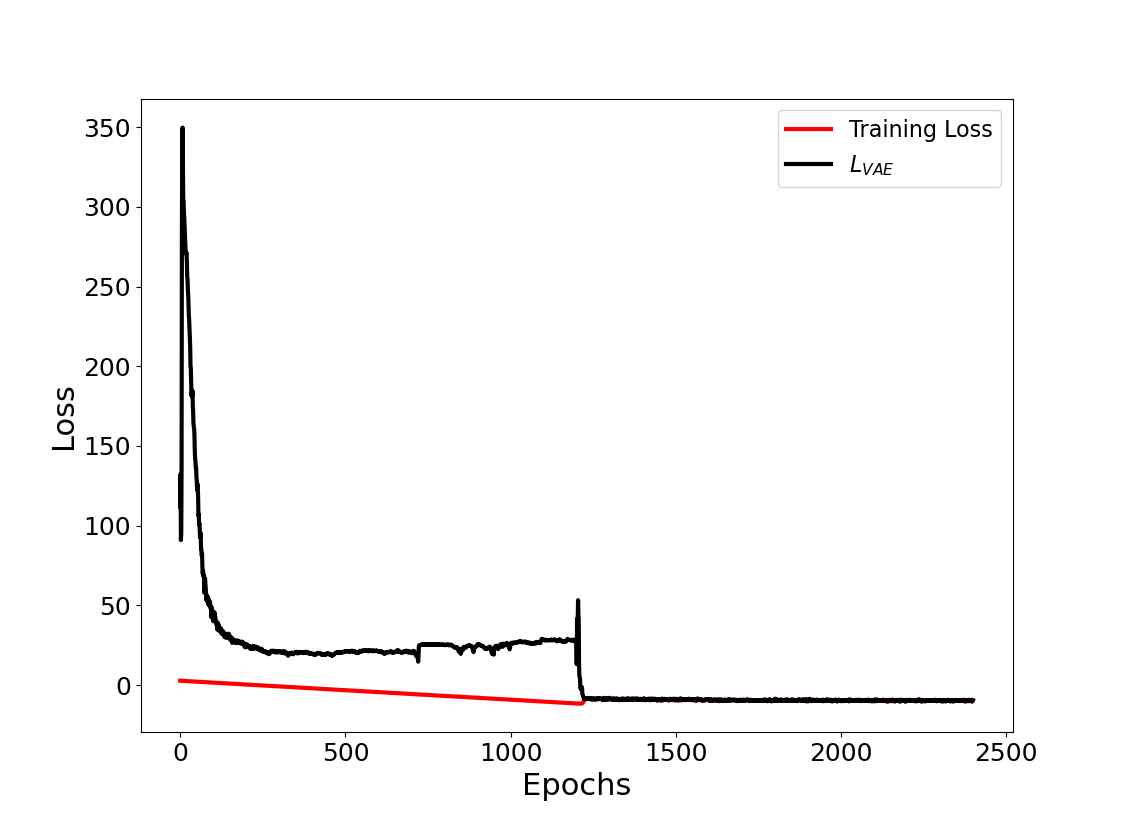}
    \includegraphics[width=.49\textwidth,angle=0,clip,trim=0pt 0pt 0pt 0pt]{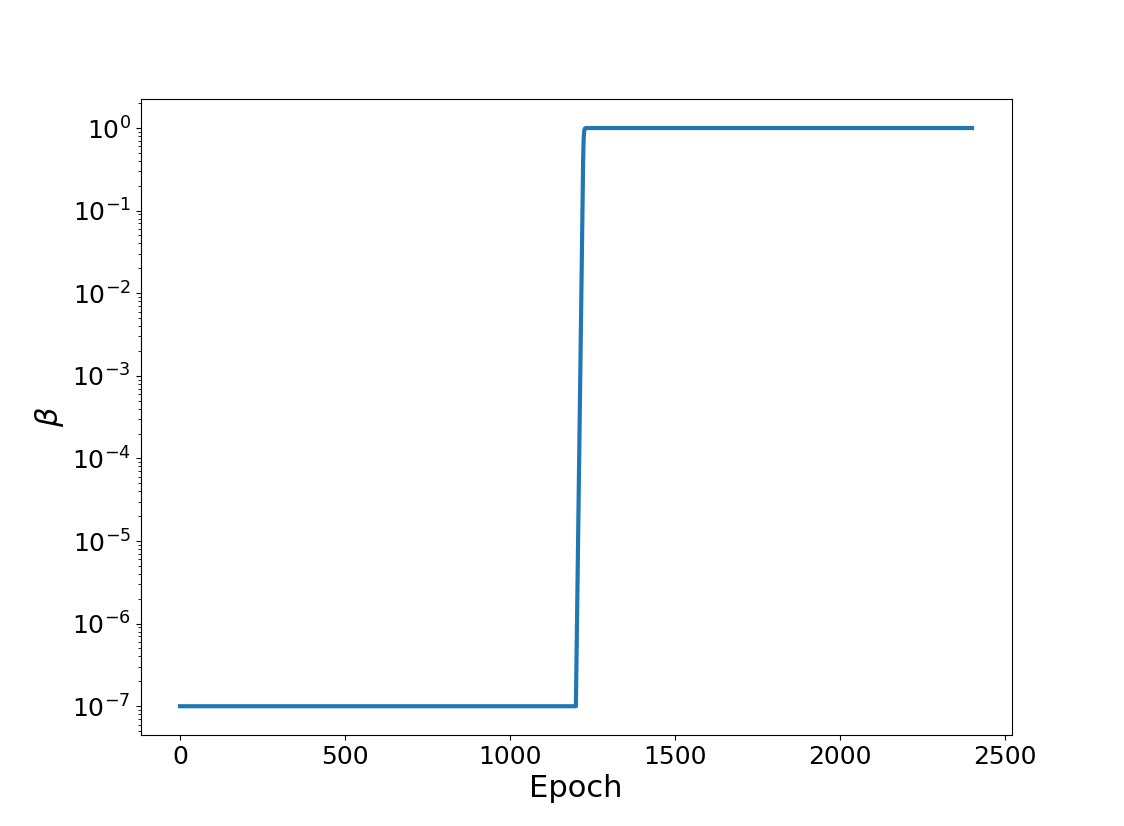}
    \caption{Training with initial increased weight on reconstruction loss helps to avoid over-regularized local minima.}
    \label{fig:training_over_reg}
\end{figure}

To illustrate the avoidance of these over-regularized local minima using our training method, Figure \ref{fig:training_over_reg} shows the training losses and $\beta$ as a function of epoch. The VAE loss reaches a local minimum but continues to increase as the true training loss decreases. With a small initial $\beta$ ($10^{-7}$), great emphasis is placed on the reconstruction loss. When $\beta$ begins to increase, the VAE is 'past' the over-regularized region and the training loss rapidly converges to the VAE loss, obtaining a desirable solution. 

\begin{figure}[h!]
    \centering
    \captionsetup{width=.8\linewidth}
    \includegraphics[width=.49\textwidth,angle=0,clip,trim=0pt 0pt 0pt 0pt]{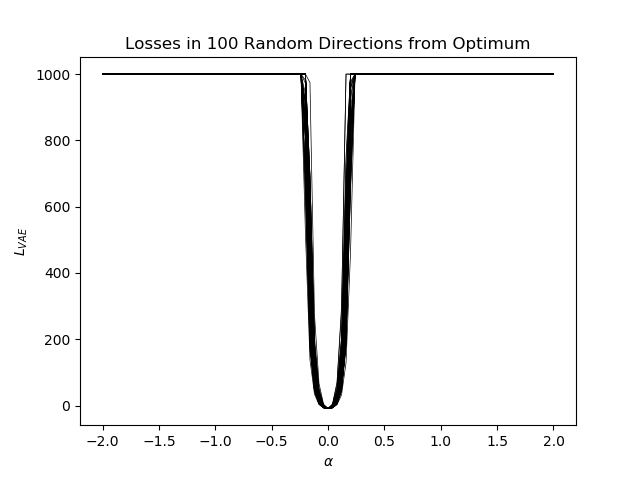}
    \caption{Loss along 100 (of 1,000) random lines emanating from a desirable solution of the DenseVAE architecture. Note that the loss is limited to 1,000 for illustration purposes}
    \label{fig:random_loss_directions_dense}
\end{figure}

Of interest is that this over-regularized local minima region does not fully surround the 'desirable' region. Instead of interpolating in parameter space between random initializations and a converged solution, lines away from the converged solution along 1,000 random directions in parameter space are created and the loss plotted along each. Figure \ref{fig:random_loss_directions_dense} illustrates that indeed there  no local minima are found around the converged solution. We note that there are around 800,000 training parameters in this case, so 1,000 random directions may not completely encapsulate the loss landscape around this solution. 

The Xavier uniform weight initialization scheme, and most other initialization schemes, limit the norm of the parameters in parameter space to near the origin. The local minima region exists only between the converged solution region and points in parameter space near the origin. In this case, there may be alternative initialization schemes which can greatly aid in the convergence of VAEs. This has been observed in \cite{sitzmann2020implicit} where the initialization scheme proposed greatly accelerates the speed of convergence \emph{and} accuracy of reconstruction. 

\begin{figure}[h!]
    \centering
    \captionsetup{width=.8\linewidth}
    \includegraphics[width=.49\textwidth,angle=0,clip,trim=0pt 0pt 0pt 0pt]{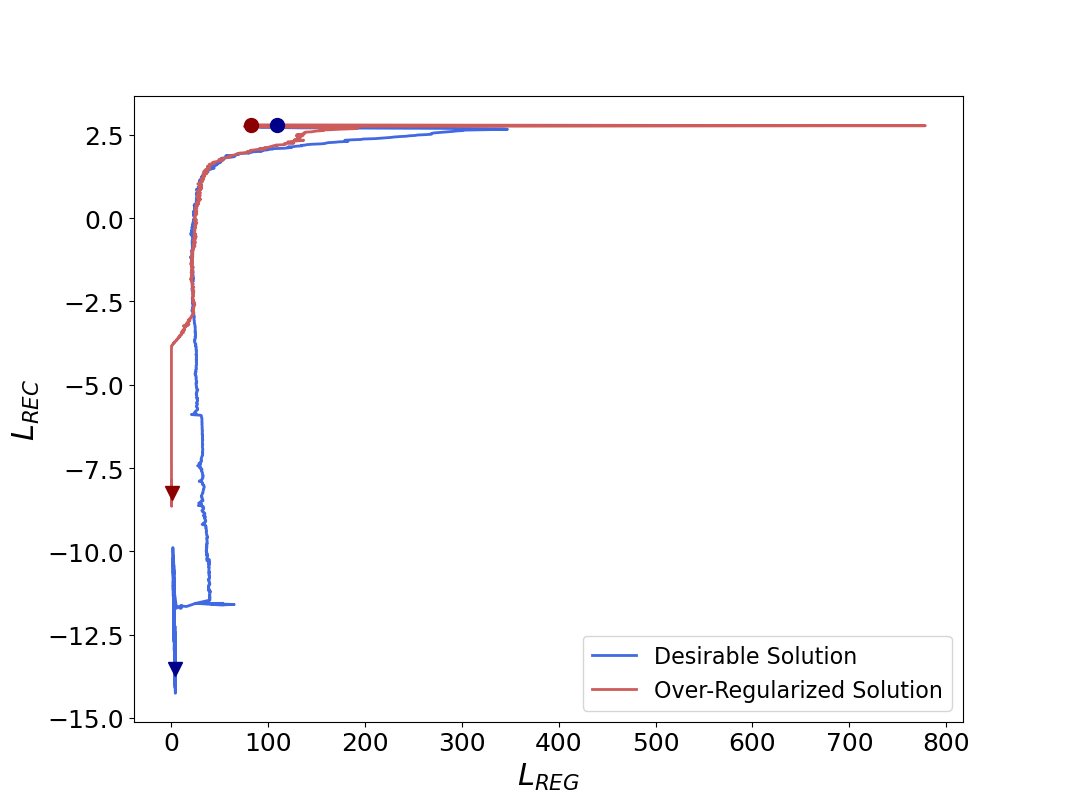}
    \includegraphics[width=.49\textwidth,angle=0,clip,trim=0pt 0pt 0pt 0pt]{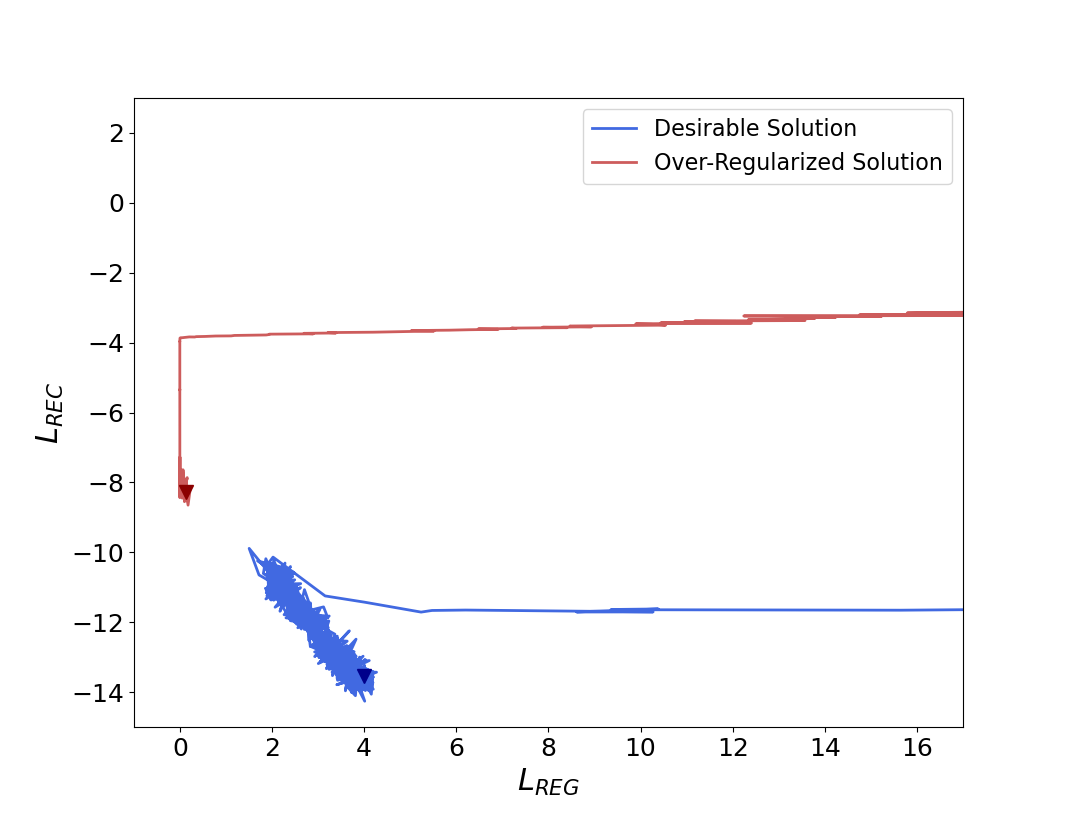}
    \caption{RD plane illustrating training convergence of both desirable and over-regularized solutions to the RD curve. (\emph{right}) Scale adjusted.}
    \label{fig:over_reg_RD_curve}
\end{figure}

Over-regularized local minima follow a similar path during training as desirable solutions. A region of attraction exists in the loss landscape, and falling too close to this region will result in an over-regularized solution, illustrated in Figure \ref{fig:over_reg_RD_curve}. One VAE which obtains a desirable solution shares a similar initial path with an over-regularized solution. Plotted are the VAE losses computed during training, not the training losses. The over-regularized solution breaks from the desired path too early, indicating a necessity for a longer reconstruction-heavy phase. 

\begin{figure}[h!]
    \centering
    \captionsetup{width=.8\linewidth}
    \includegraphics[width=.49\textwidth,angle=0,clip,trim=0pt 0pt 0pt 0pt]{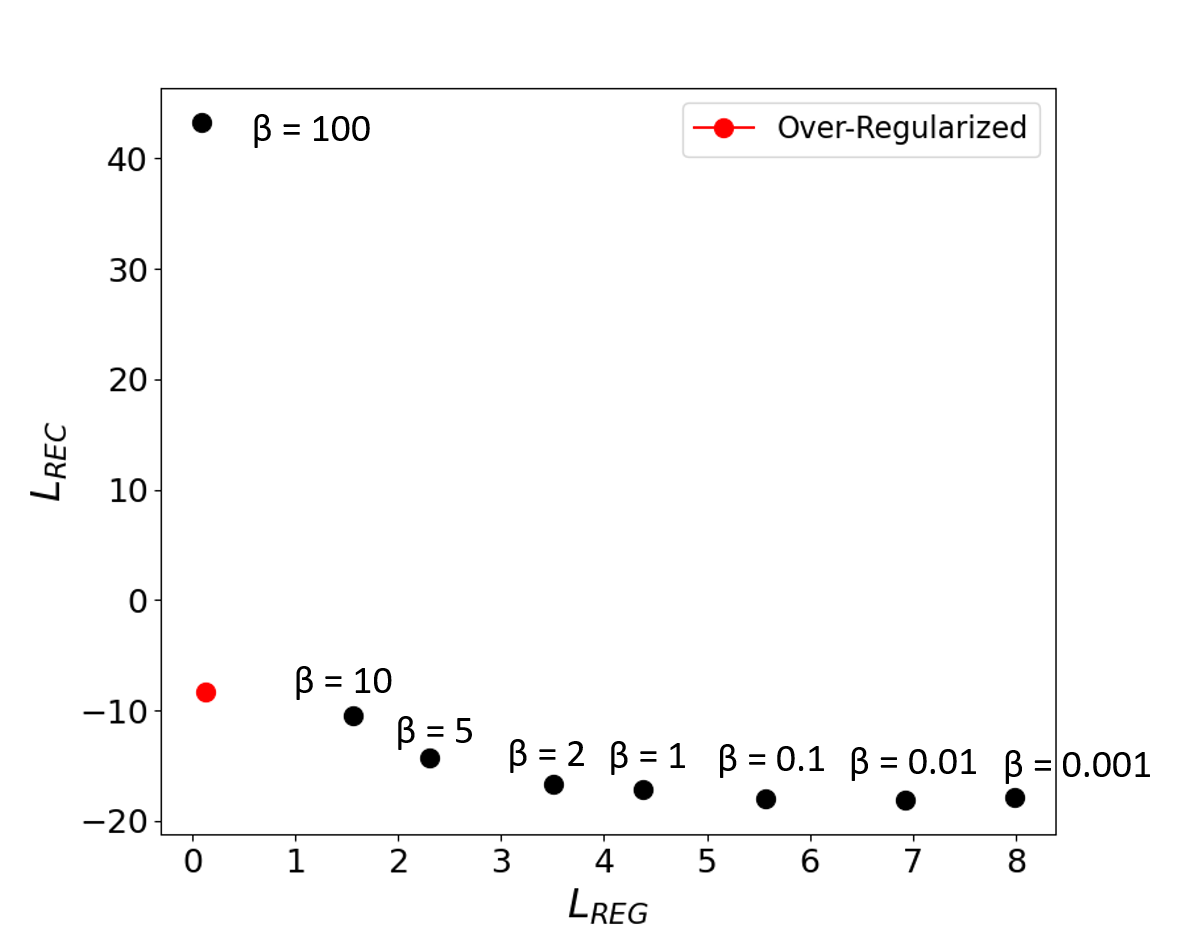}
    \caption{RD plane with points corresponding (from left to right) to $\beta$ = [100, 10, 5, 2, 1, 0.1, 0.01, 0.001]. Many values of $\beta$ between 5 and 100 fall into the over-regularized solution.}
    \label{fig:VAE_RD_curve}
\end{figure}

Training many VAEs with various $\beta$ values facilitates a visualization of over-regularization in the RD plane. Each point in Figure \ref{fig:VAE_RD_curve} shows the loss values of converged VAEs trained with different values of $\beta$. The over-regularized region of attraction prevents convergence to desirable solutions for many values of $\beta$. Interpolating in parameter space between each of these points (corresponding to a VAE with its own converged parameters) using the base VAE loss ($\beta = 1$), no other points on the RD curve are local minima of the VAE loss. In Figure \ref{fig:over_reg_RD_curve}, we observe that during training, the desirable solution reaches the RD curve but continues toward the final solution.

\subsection{Properties of Desirable Solutions}\label{sec:varying_properties}
\begin{figure}[h!]
    \centering
    \captionsetup{width=.8\linewidth}
    \includegraphics[width=.49\textwidth,angle=0,clip,trim=0pt 0pt 0pt 0pt]{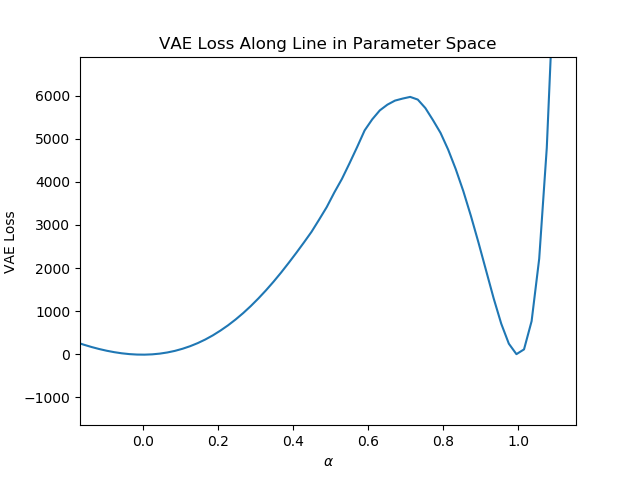}
    \includegraphics[width=.49\textwidth,angle=0,clip,trim=0pt 0pt 0pt 0pt]{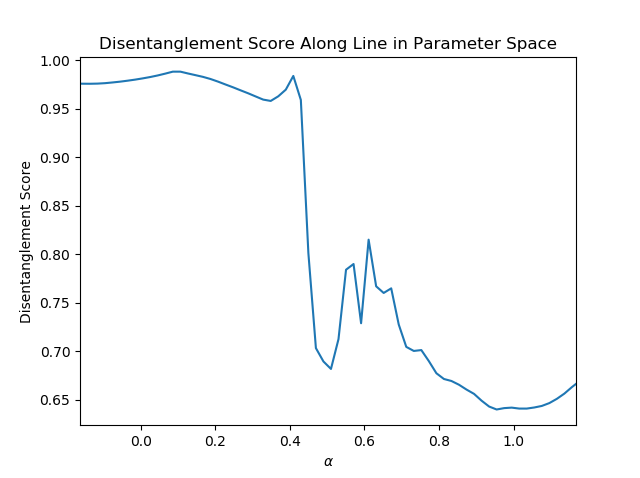}
    \caption{\emph{left} Loss variation along a line in parameter space between two converged solutions containing identical hyperparameters and training method but different network parameter initializations. \emph{right} Disentanglement score along the same line.}
    \label{fig:loss_scape_comp_1D}
\end{figure}

Avoiding over-regularization aids in convergence to solutions characterized by low reconstruction error. Among solutions with similar final loss values, inconsistencies remain in latent properties. Two identical VAEs initialized separately often converge to similar loss values, but one may exhibit disentanglement while the other does not. This phenomenon is also explored in \cite{locatello2019challenging} and \cite{rolinek2019variational}. Two VAEs are trained with identical architectures, hyperparameters, and training method; they differ only in the random initialization of network parameters $P$. 
We denote the optimal network parameters found from one initialization as $P_1$ and optimal network parameters found from a separate initialization $P_2$. The losses for each converged solution are quite similar ($L_{VAE_1} \approx -9.50, \;\; L_{VAE_2} \approx -9.42$); however, disentanglement properties of each are dramatically different. We interpolate between these two solutions in parameter space (Eq \ref{eq:alpha_line}) and record losses and disentanglement scores along the line (Figure \ref{fig:loss_scape_comp_1D}). The first network contains a nearly perfectly disentangled latent representation while the second network does not produce a disentangled representation. It is evident that multiple local minima exist in parameter space which converge to similar values in the loss landscape, but contain very different latent correlations. Local minima exist throughout the loss landscape, and with each initialization, a different local minimum may be found. Many such differing solutions are found throughout our experiments. This phenomenon is partially due to invariance of the ELBO to rotations of the latent space when using rotationally invariant priors. Disentanglement is heavily dependent on a factorized representation of the latent representation. With rotations not affecting the training loss, learning a disentangled representation seems to be somewhat random in this case. 

This phenomenon exhibits the difficulties in disentangling generative parameters in an unsupervised manner; without prior knowledge of the factors of variation, conclusions cannot be drawn regarding disentanglement by observing loss values alone. In controlled experiments, knowledge of the underlying factors of variation is available, but when only data is available, full knowledge of such factors is often not. It is encouraging that the VAE does have the power to disentangle generative parameters in an unsupervised setting, but the nature of disentanglement must first be understood better to create identifying criterion. 
\section{Characterizing Disentanglement} \label{sec:KLE2}
In this section, we explore the relationship between disentanglement, the aggregated posterior ($q_\phi(z)$), and the generative parameter distribution ($p(\theta)$) by incrementally increasing the complexity of $p(\theta)$. Disentanglement is first illustrated to be achievable but difficult using the classic VAE assumptions and loss due to a lack of enforcement of the rotation of the latent space caused by rotationally-invariant priors. Hierarchical priors are shown to aid greatly in disentangling the latent space by learning non-rotationally-invariant priors which enforce a particular rotation of the latent space through the regularization loss. 

\subsection{Standard Normal Generative Distributions}

The intrinsic dimensionality of the data is set to $p=2$ with a generative parameter distribution $p(\theta) = \mathcal{N}(\theta;0, I_{2\times 2})$, the standard normal distribution. Limiting $p$ to 2 aids greatly in the visualization of the latent space and understanding of the ideas investigated. The standard latent prior is identical to the generative parameter in this case, creating a relatively simple problem for the VAE. 

\begin{figure}[h!]
    \centering
    \captionsetup{width=.8\linewidth}
    \includegraphics[width=.49\textwidth,angle=0,clip,trim=0pt 0pt 0pt 0pt]{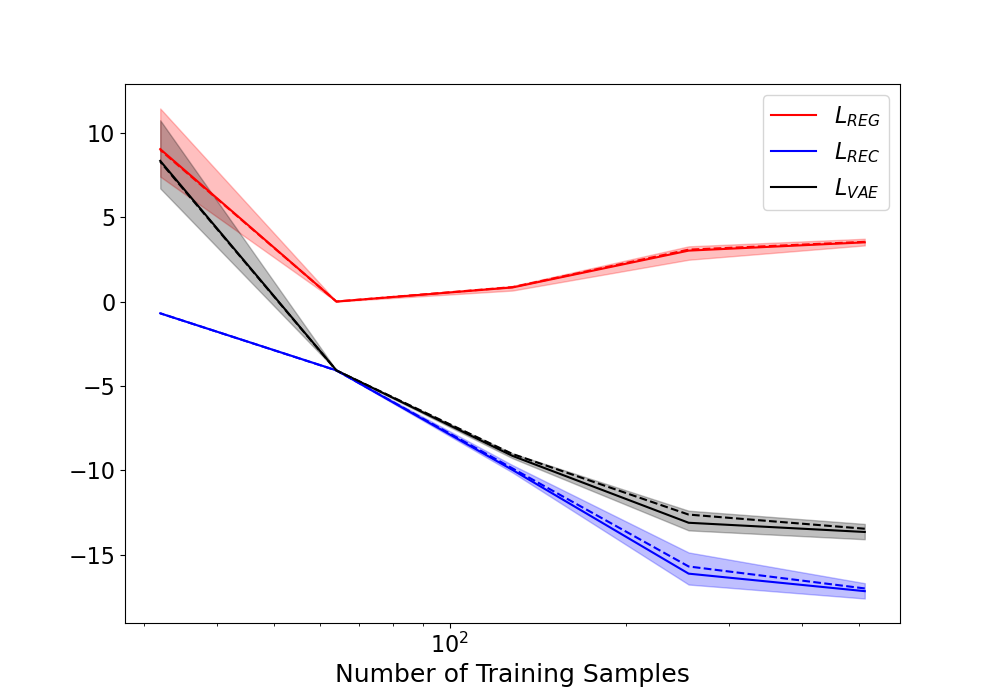}
    \includegraphics[width=.49\textwidth,angle=0,clip,trim=0pt 0pt 0pt 0pt]{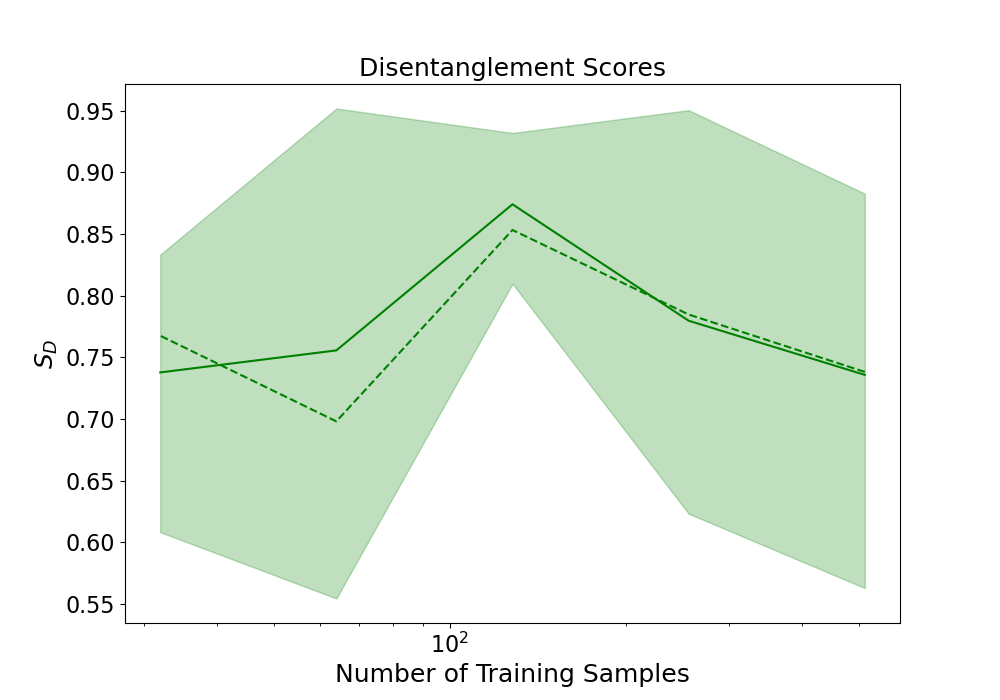}
    \caption{Solid lines indicate averages over training data for 10 VAEs trained at each point. Dashed lines represent averages over testing data. Ranges indicate minimum and maximum values. \emph{left} Converged VAE losses for various numbers of training samples. \emph{right} Converged VAE disentanglement score as a function of number of training samples.}
    \label{fig:num_train_study}
\end{figure}

Using the architecture described in Section \ref{sec:architecture}, the relationship between regularization, reconstruction, and disentanglement and the number of training samples is illustrated in Figure \ref{fig:num_train_study}. Using the training method detailed in Section \ref{sec:issues}, all results are obtained with $\beta$ = 1 during training and a latent dimension $n = 2$. For every number of training data ([32, 64, 128, 256, 512]), 10 VAEs are trained. A similar study is performed in \cite{locatello2019challenging} with a greater sample size. Reconstruction losses continue to fall with the number of training data, indicating improved reconstruction of the data with increased number of samples; however, the regularization loss increases slightly with the number of training data. With too few samples, reconstruction performance is very poor and over-regularization (near zero regularization loss) seems unavoidable. Clear and consistent correlations exist among the loss values and number of training data, but disentanglement properties vary greatly among converged VAEs (Section \ref{sec:varying_properties}). The compressed representations range from nearly perfect disentanglement to nearly completely entangled. 

\begin{figure}[h!]
    \centering
    \captionsetup{width=.8\linewidth}
    \includegraphics[width=.99\textwidth,angle=0,clip,trim=0pt 0pt 0pt 0pt]{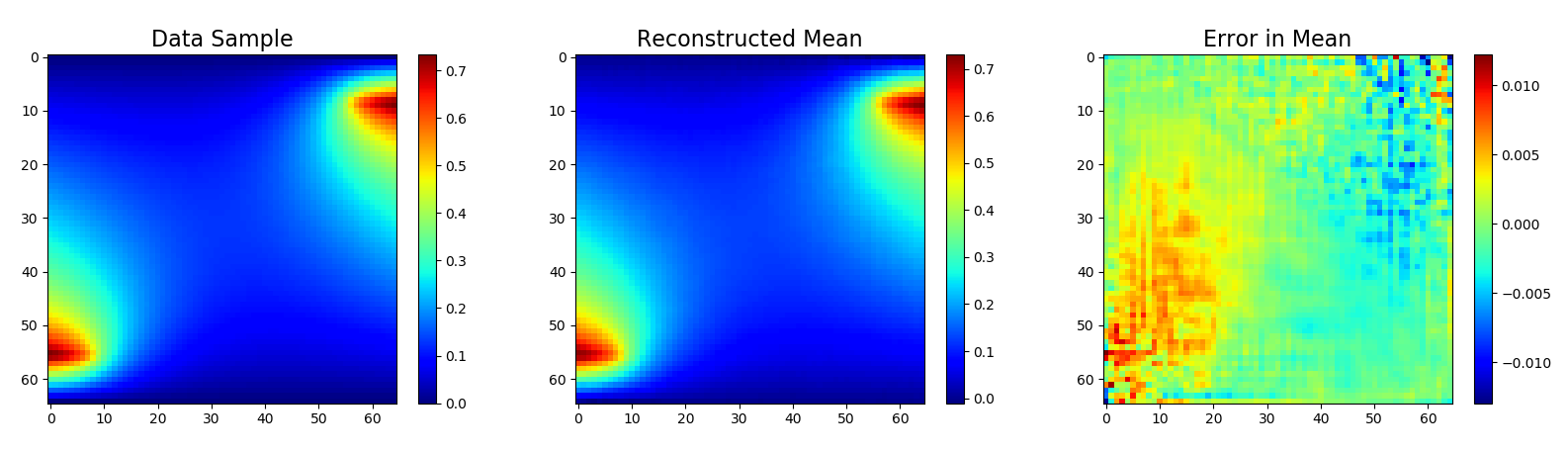}
    \caption{(\emph{left}) Data sample from unseen testing dataset. (\emph{center}) Reconstructed data sample from trained VAE. (\emph{right}) Error in the reconstruction mean.}
    \label{fig:VAE_results}
\end{figure}

\begin{figure}[h!]
    \centering
    \captionsetup{width=.8\linewidth}
    \includegraphics[width=.45\textwidth,angle=0,clip,trim=0pt 0pt 0pt 0pt]{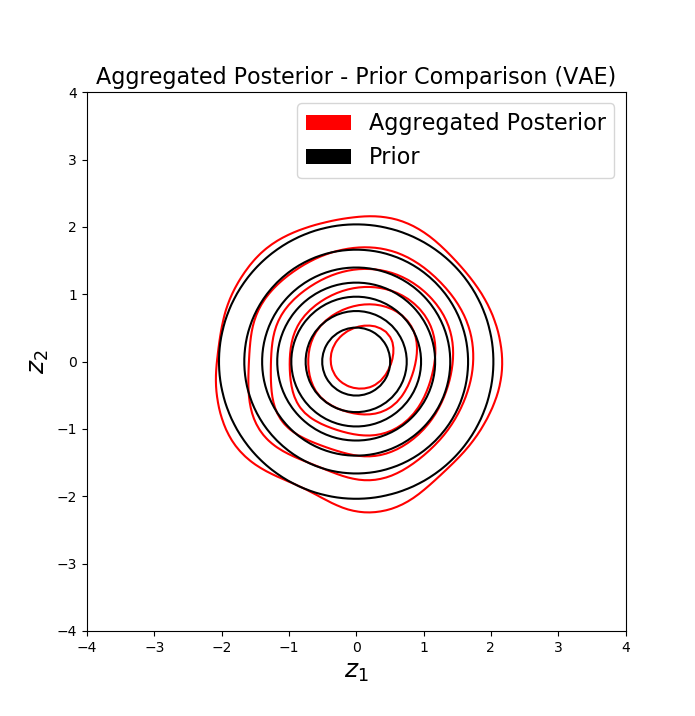}
    \caption{Comparison of aggregated posterior ($p_\phi(z)$) and prior ($p(z)$) distributions.}
    \label{fig:VAE_agg_post_comp}
\end{figure}

\begin{figure}[h!]
    \centering
    \captionsetup{width=.8\linewidth}
    \includegraphics[width=.75\textwidth,angle=0,clip,trim=0pt 0pt 0pt 0pt]{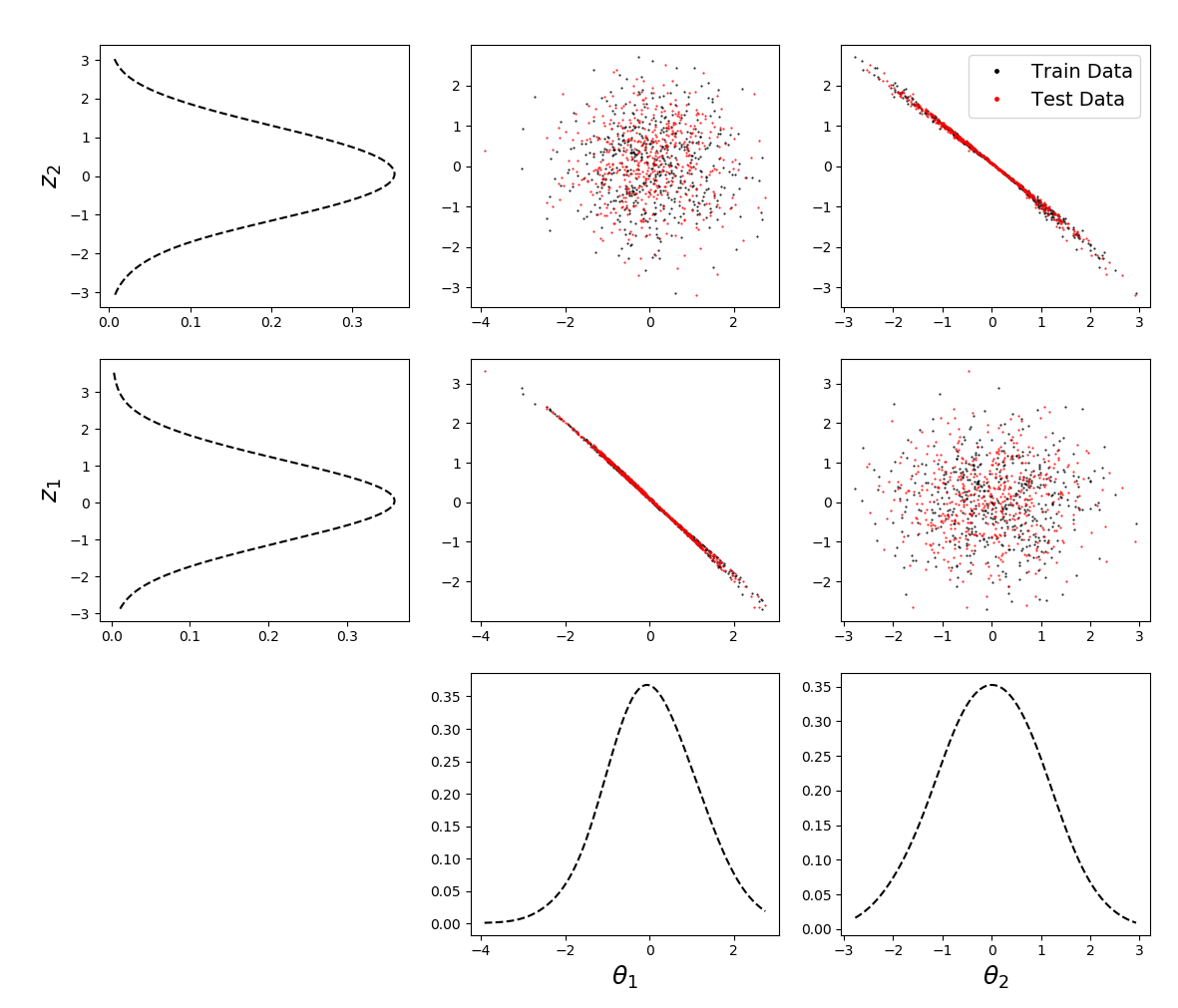}
    \includegraphics[width=.7\textwidth,angle=0,clip,trim=0pt 0pt 0pt 0pt]{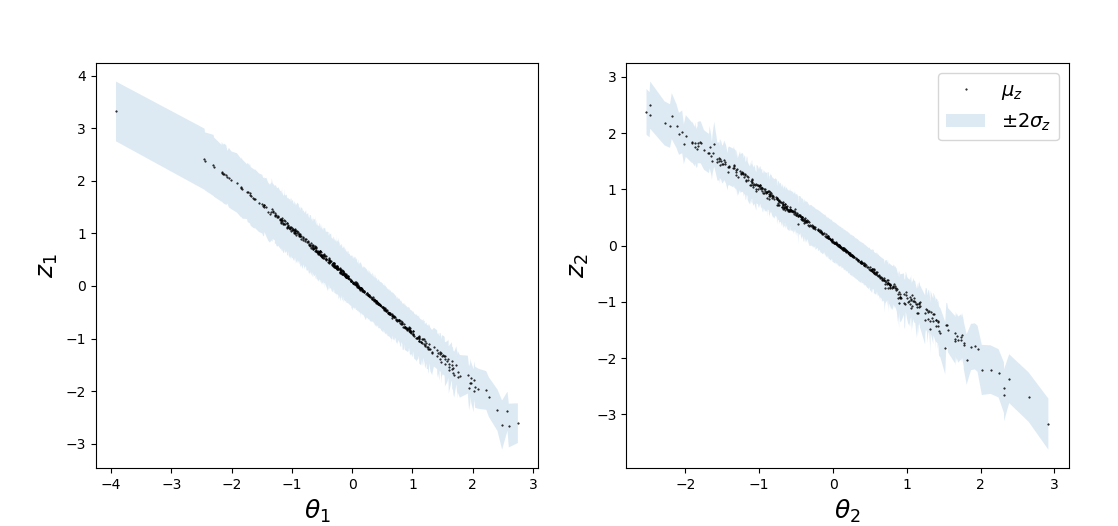}
    \caption{(\emph{upper}) Correlations between dimensions of generative parameters and mean of latent parameters. Also shown are the empirical marginal distributions of each parameter. (\emph{lower}) Correlations between generative parameters and latent parameters with uncertainty for test data only.}
    \label{fig:VAE_latent_correlations}
\end{figure}

Although disentanglement properties are inconsistent between experiments, desirable properties of disentanglement are often observed. Training is performed using the maximum amount of available data (512 snapshots), and analysis included for 512 testing samples on the KLE2 dataset (regardless of $p(\theta)$). Regularization loss is large during the reconstruction phase in which $\beta_0 = 10^{-7}$, and the y-axis is truncated for clarity. A comparison between a test data sample and the reconstructed mean using the trained VAE is depicted in Figure \ref{fig:VAE_results}, showing little error between the mean $\mu_\psi(z)$ of the decoding distribution and the input data sample. With small reconstruction error, a disentangled latent representation is learned. Figure \ref{fig:VAE_agg_post_comp} illustrates the aggregated posterior matching the prior distribution in shape. This is unsurprising with a generative parameter and prior distribution match and an expressive network architecture. Finally, Figure \ref{fig:VAE_latent_correlations} shows the correlation between the generative parameters of the training and testing data against the latent distribution as a qualitative measure of disentanglement. Each latent dimension is tightly correlated to a single but different generative parameter. Figure \ref{fig:VAE_latent_correlations} also illustrates the uncertainty in the latent parameters, effectively $q_\phi(z|\theta)$. The latent representation is fully disentangled; each latent parameter contains only information about a single generative factor.

\subsection{Non Standard Gaussian Generative Distributions}

\begin{figure}[h!]
    \centering
    \captionsetup{width=.8\linewidth}
    \includegraphics[width=.99\textwidth,angle=0,clip,trim=0pt 0pt 0pt 0pt]{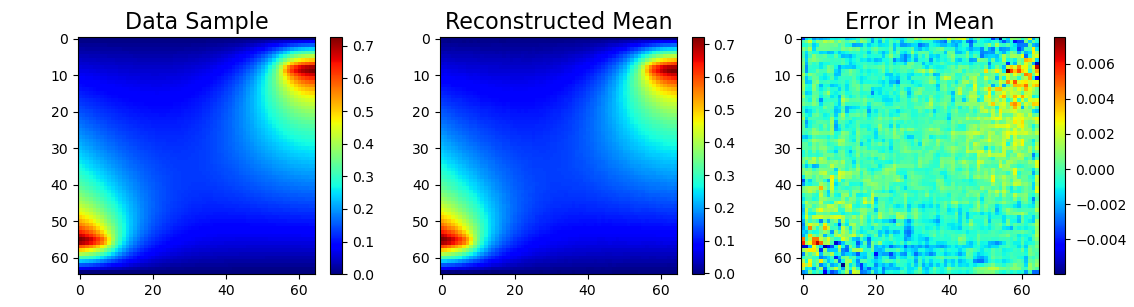}
    \includegraphics[width=.97\textwidth,angle=0,clip,trim=0pt 0pt 0pt 0pt]{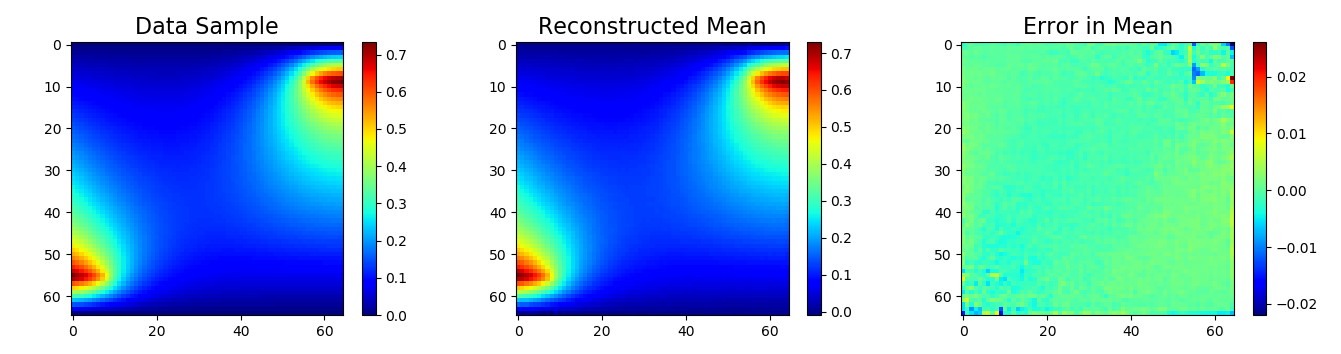}
    \caption{(\emph{top}) Reconstruction accuracy of a test sample on trained VAE without hierarchical network, (\emph{bottom}) with hierarchical network.}
    \label{fig:kle2_trans_gauss_recon}
\end{figure}
\begin{figure}[h!]
    \centering
    \captionsetup{width=.8\linewidth}
    \includegraphics[width=.49\textwidth,angle=0,clip,trim=0pt 0pt 0pt 0pt]{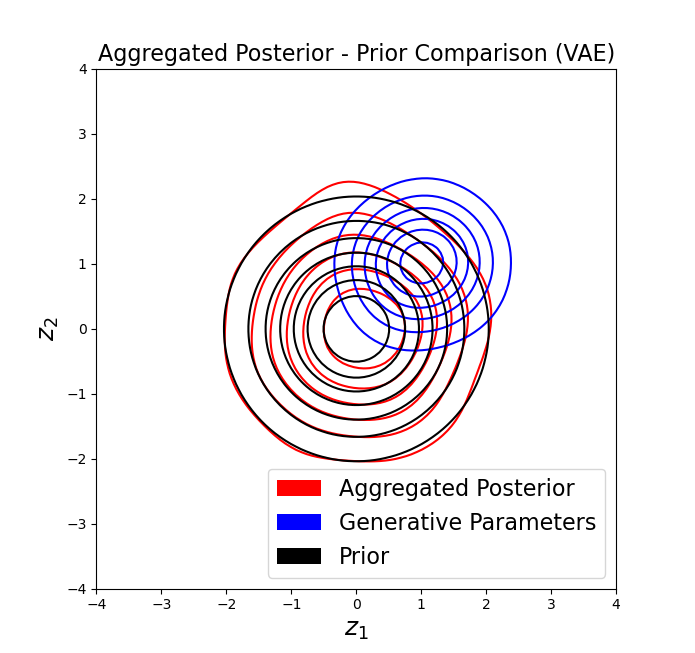}
    \includegraphics[width=.49\textwidth,angle=0,clip,trim=0pt 0pt 0pt 0pt]{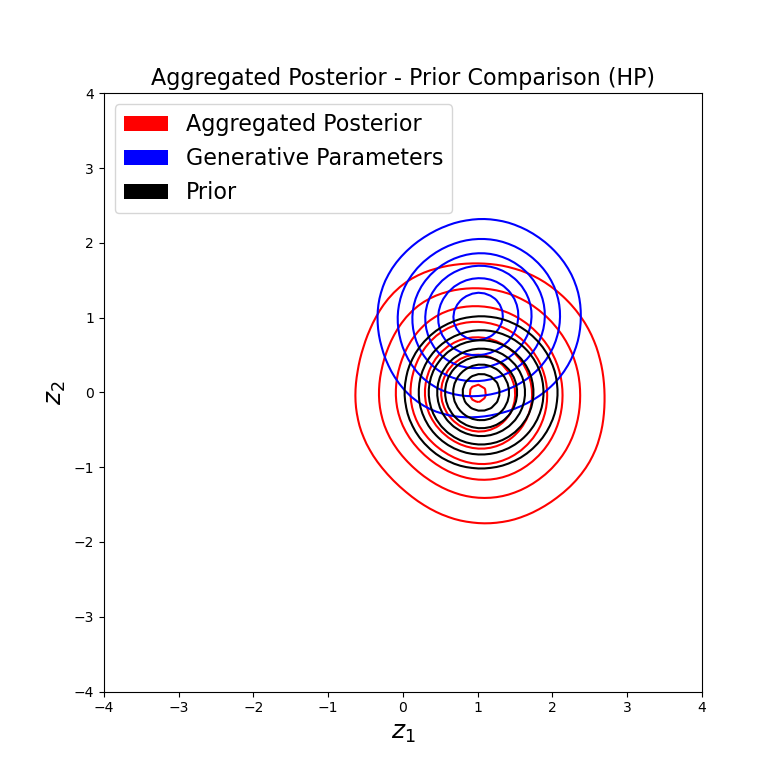}
    \caption{(\emph{left}) Aggregated poster, prior, and generative parameter distribution comparison on VAE without hierarchical network, (\emph{right}) with hierarchical network.}
    \label{fig:kle2_trans_gauss_agg_post}
\end{figure}

The generative parameter distribution and the prior are identical (independent standard normal) in the previous example. Most often, however, knowledge of the generative parameters is not possessed. The specified prior in this case is unlikely to match the generative parameter distribution. The next example illustrates the application of a VAE in which the generative parameter distribution and prior do not match. Another KLE2 dataset is generated with a non standard Gaussian generative parameter distribution. The generative parameter distribution is Gaussian, but scaled and translated relative to the previous example $p(\theta) = \mathcal{N}(\theta; [1;1], [0.5, 0; 0, 0.5])$.

\begin{figure}[h!]
    \centering
    \captionsetup{width=.8\linewidth}
    \includegraphics[width=.49\textwidth,angle=0,clip,trim=0pt 0pt 0pt 0pt]{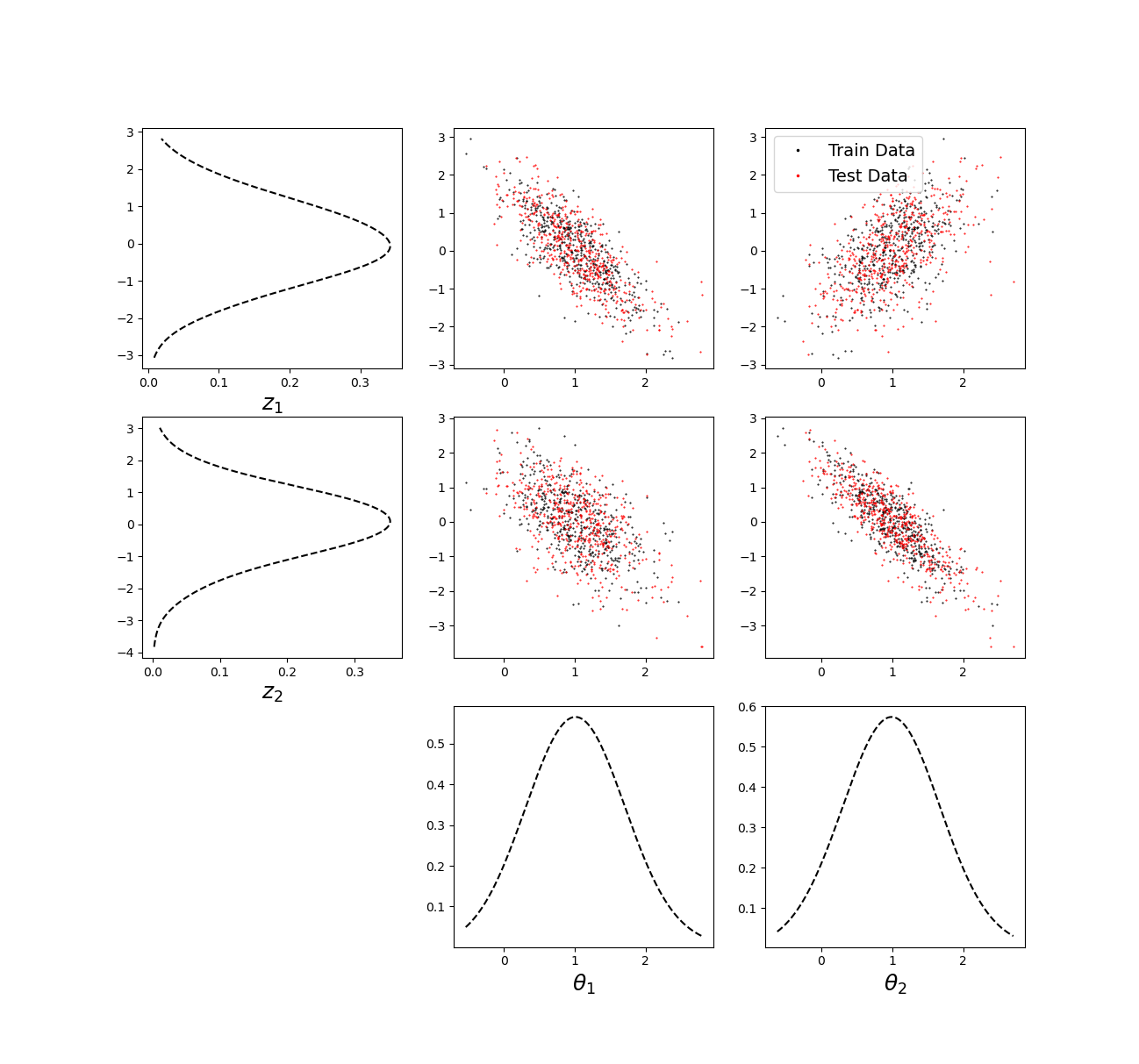}
    \includegraphics[width=.49\textwidth,angle=0,clip,trim=0pt 0pt 0pt 0pt]{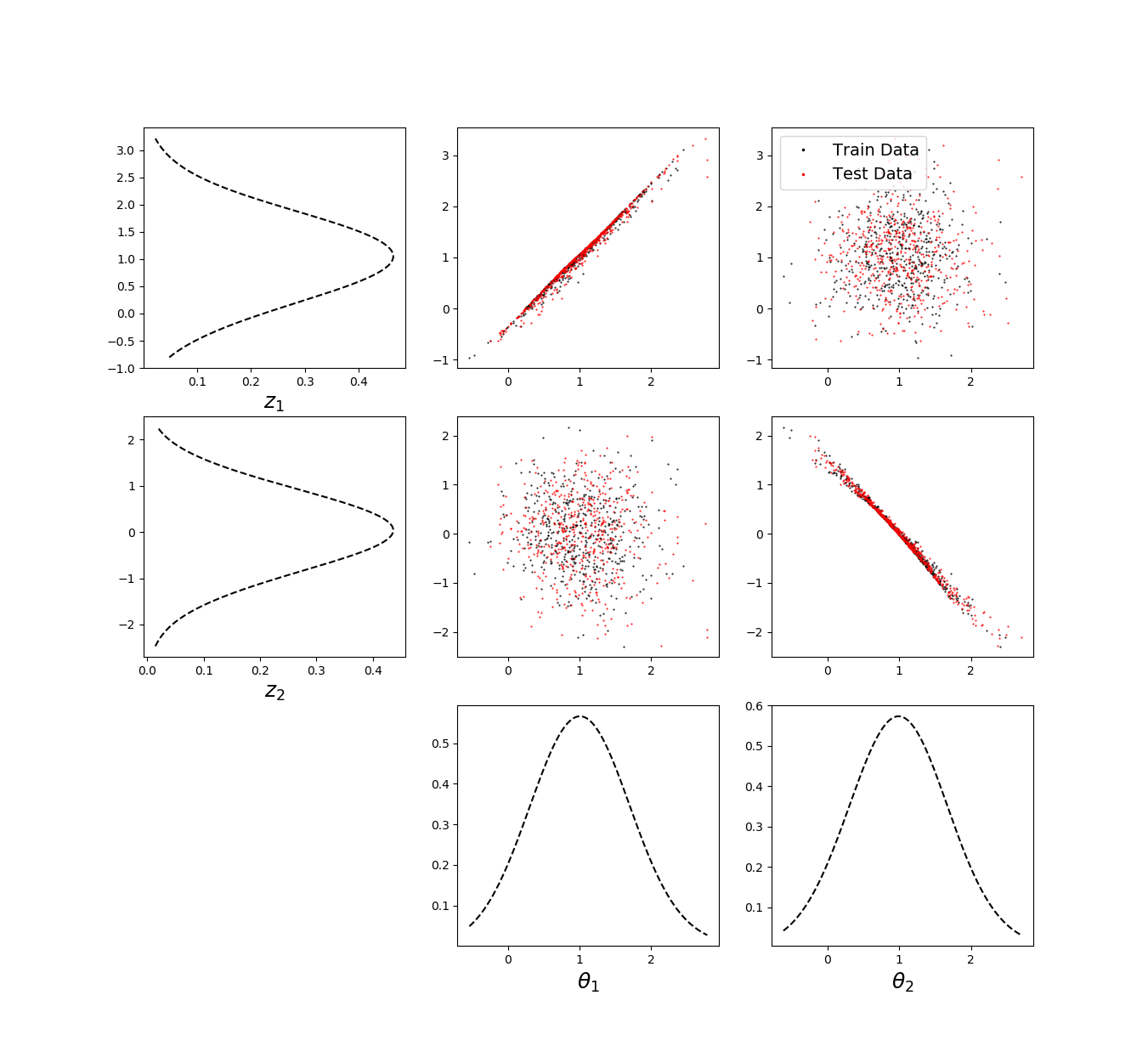}
    \caption{(\emph{left}) Qualitative disentanglement in VAE trained without hierarchical network, (\emph{right}) with hierarchical network.}
    \label{fig:kle2_trans_gauss_dis}
\end{figure}

Training a standard VAE on this dataset results in high reconstruction accuracy, but undesirable disentanglement after many trials. With the use of an additional hierarchical prior network, good disentanglement can be achieved even with a mismatch in the prior and generative parameter distributions. The sub-prior (see Section \ref{sec:hierarchical_priors}) is the standard normal distribution, but the hierarchical network learns a non-standard normal prior. Still, the learned prior and generative parameter distributions do not match. Figures \ref{fig:kle2_trans_gauss_recon}, \ref{fig:kle2_trans_gauss_agg_post}, and \ref{fig:kle2_trans_gauss_dis} illustrate comparisons in results obtain from the VAE with and without the hierarchical prior network. When using hierarchical priors, the learned prior and aggregated posterior match reasonably well but do not match the generative parameter distribution. However, this does not matter as long as the latent representation is not \emph{rotated} relative to the generative parameter distribution, as illustrated in the next example. Low reconstruction error and disentanglement are observed using hierarchical priors, but disentanglement was never observed using the standard VAE after many experiments. This may be because $\beta$ is not large enough to enforce a regularization loss large enough to produced an aggregated posterior aligned with the axes of the generative parameter distribution. Therefore, the rotation of the learned latent representation will be random and disentanglement is unlikely to be observed, even in two dimensions. The hierarchical network consistently enforces a factorized aggregated posterior, which is essential for disentanglement when generative parameters are independent. One potential cause of this is the learning of non-rotationally-invariant priors, such as a factorized Gaussian with independent scaling in each dimension. The ELBO loss in this case \emph{is} affected by rotations of the latent space, aligning the latent representations to the axes of the generative parameters.

\begin{figure}[h!]
    \centering
    \captionsetup{width=.8\linewidth}
    \includegraphics[width=.49\textwidth,angle=0,clip,trim=0pt 0pt 0pt 0pt]{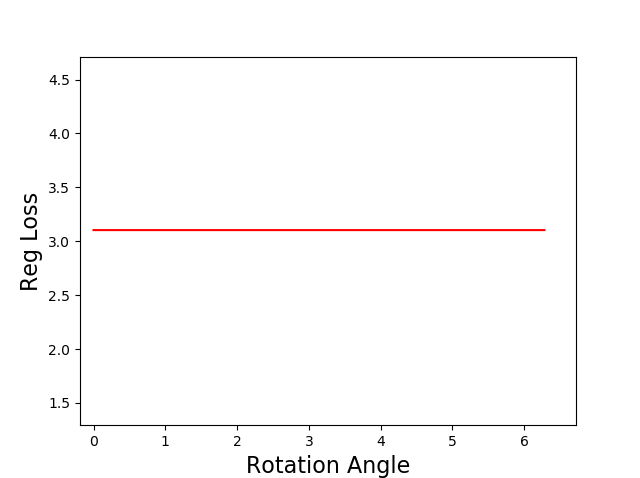}
    \includegraphics[width=.49\textwidth,angle=0,clip,trim=0pt 0pt 0pt 0pt]{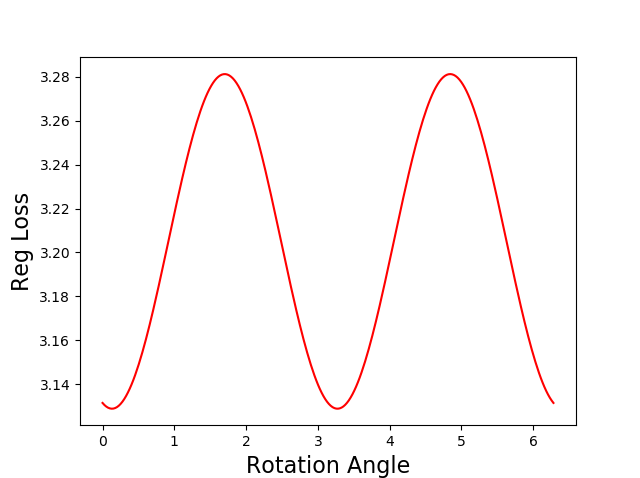}
    \caption{(\emph{left}) Regularization loss unaffected by latent rotation when training with rotationally-invariant priors, (\emph{right}) regularization loss is affected by latent rotation when training with non-rotationally-invariant priors.}
    \label{fig:latent_rot}
\end{figure}

A latent rotation can be introduced such that the reconstruction loss is unaffected, but regularization loss changes with rotation. Introducing a rotation matrix $A$ with angle of rotation $\omega$ to rotate the latent distribution, the encoding distribution becomes $q_\phi(z|y) = \mathcal{N}(z; A\mu_\phi(y), A\textrm{diag}(\sigma_\phi(y))A^T)$. Reversing this rotation when computing the decoding distribution (i.e. $p_\psi(y|z) = \mathcal{N}(y; \mu_\phi(A^Tz), \textrm{diag}(\sigma_\psi(A^Tz)))$) preserves the reconstruction loss. However, the regularization loss can be plotted as a function of the rotation angle. Figure \ref{fig:latent_rot} illustrates that when a rotationally-invariant prior is used to train the VAE, regularization loss is unaffected by latent rotation. However, when the prior is non-rotationally-invariant, the regularization loss is affected by latent rotation. Thus, rotation of the latent space is enforced by the prior during training. 

Although the hierarchical prior adds some trainable parameters to the overall architecture, the increase is only $0.048\%$. This is  negligible, and it is assumed that this is not the root cause of improved disentanglement. Rather, it is the ability of the additional hierarchical network to consistently express a factorized aggregated posterior and learn non-rotationally-invariant priors which improves disentanglement. More insights are offered in the next example and Section \ref{sec:conclusion}.

\subsection{Multimodal Generative Distributions}

\begin{figure}[h!]
    \centering
    \captionsetup{width=.75\linewidth}
    \includegraphics[width=.99\textwidth,angle=0,clip,trim=0pt 0pt 0pt 0pt]{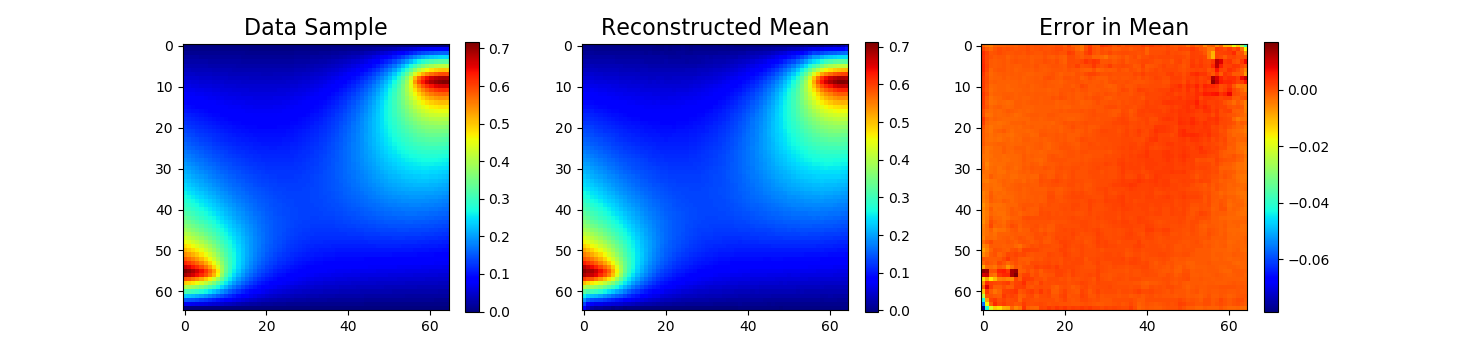}
    \includegraphics[width=.99\textwidth,angle=0,clip,trim=0pt 0pt 0pt 0pt]{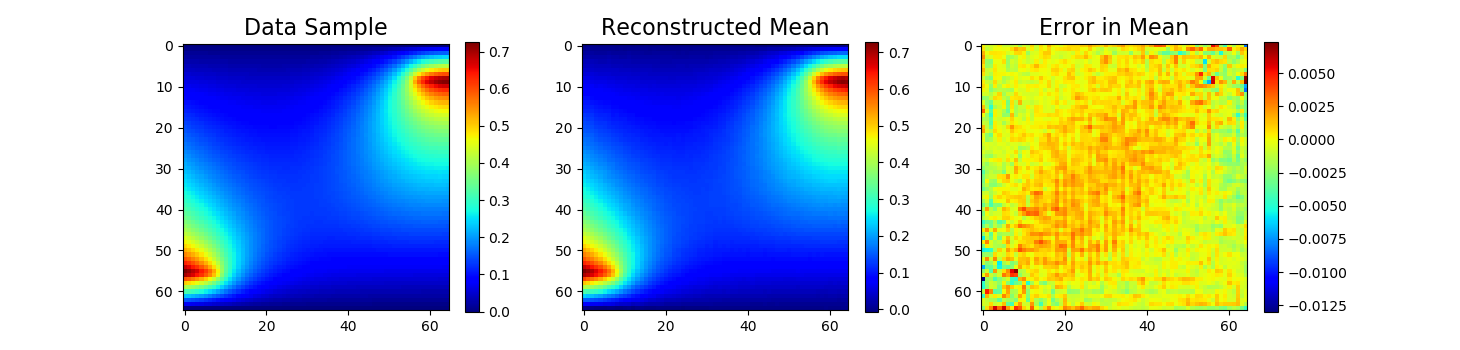}
    \caption{(\emph{top}) Reconstruction accuracy of a test sample using VAE trained on multimodal generative parameter distribution without hierarchical network, (\emph{bottom}) with hierarchical network.}
    \label{fig:kle2_bimodal_gen_2_recon}
\end{figure}

In this setup, disentanglement not only depends on a factorized $q_\phi(z)$, but the correlations in $p(\theta)$ must be preserved as well, i.e. rotations matter. The previous example illustrates a case in which the standard VAE fails in disentanglement but succeeds with the addition of hierarchical priors due to improved enforcement of a factorized $q_\phi(z)$ through learning non-rotationally-invariant priors. The generative parameter distribution is radially symmetric, thus visualization of rotations in $q_\phi(z)$ relative to $p(\theta)$ is difficult. To illustrate the benefits of using hierarchical priors for disentanglement, the final example uses data generated from a more complex generative parameter distribution with four lines of symmetry for better visualization. The generative parameter distribution is multimodal (a Gaussian mixture) and is more difficult to capture than a Gaussian distribution, but allows for better rotational visualization:

\begin{align*}
    p(\theta)  &= \frac{1}{4}\mathcal{N}(\theta; [-1;-1], [0.25, 0;0, 0.25]) + \frac{1}{4}\mathcal{N}(\theta; [1;1], [0.25, 0; 0, 0.25])\\ 
     &+\frac{1}{4}\mathcal{N}(\theta; [-1;1], [0.25, 0; 0, 0.25]) + \frac{1}{4}\mathcal{N}(\theta; [1;-1], [0.25, 0; 0, 0.25])
\end{align*}

\begin{figure}[h!]
    \centering
    \captionsetup{width=.75\linewidth}
    \includegraphics[width=.49\textwidth,angle=0,clip,trim=0pt 0pt 0pt 0pt]{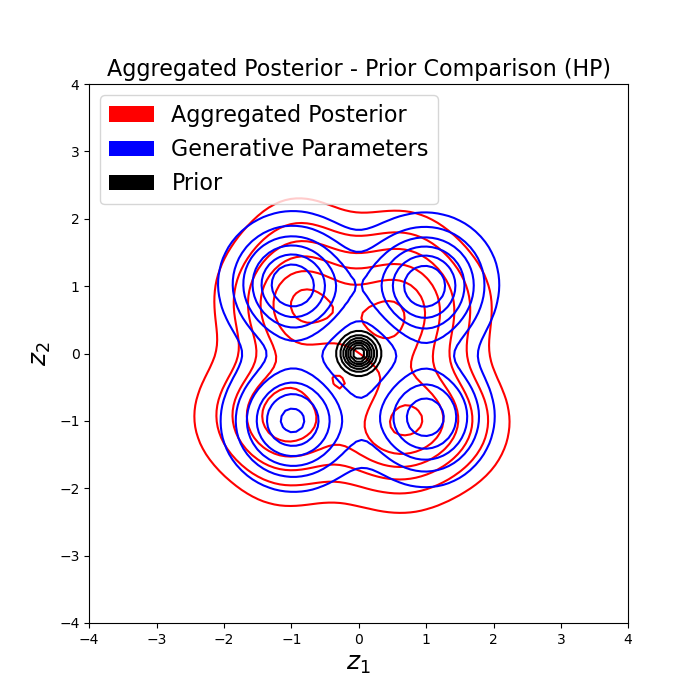}
    \includegraphics[width=.49\textwidth,angle=0,clip,trim=0pt 0pt 0pt 0pt]{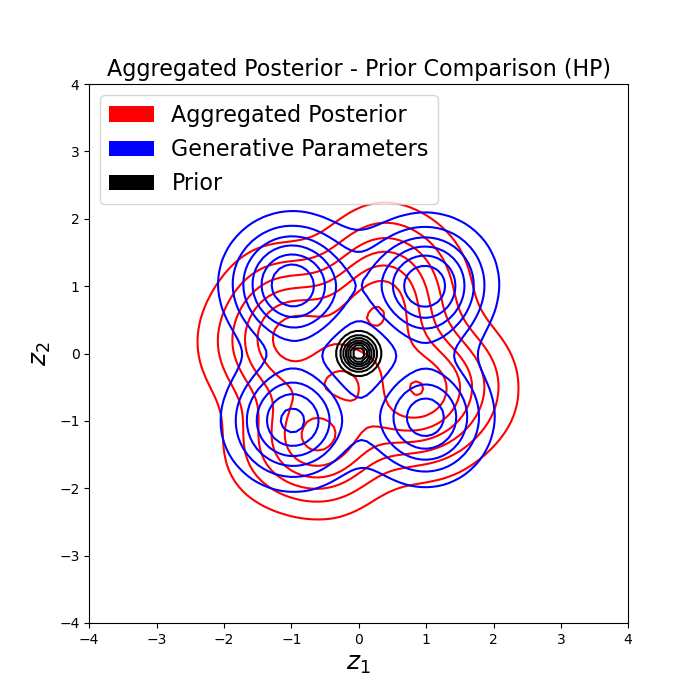}
    \includegraphics[width=.49\textwidth,angle=0,clip,trim=0pt 0pt 0pt 0pt]{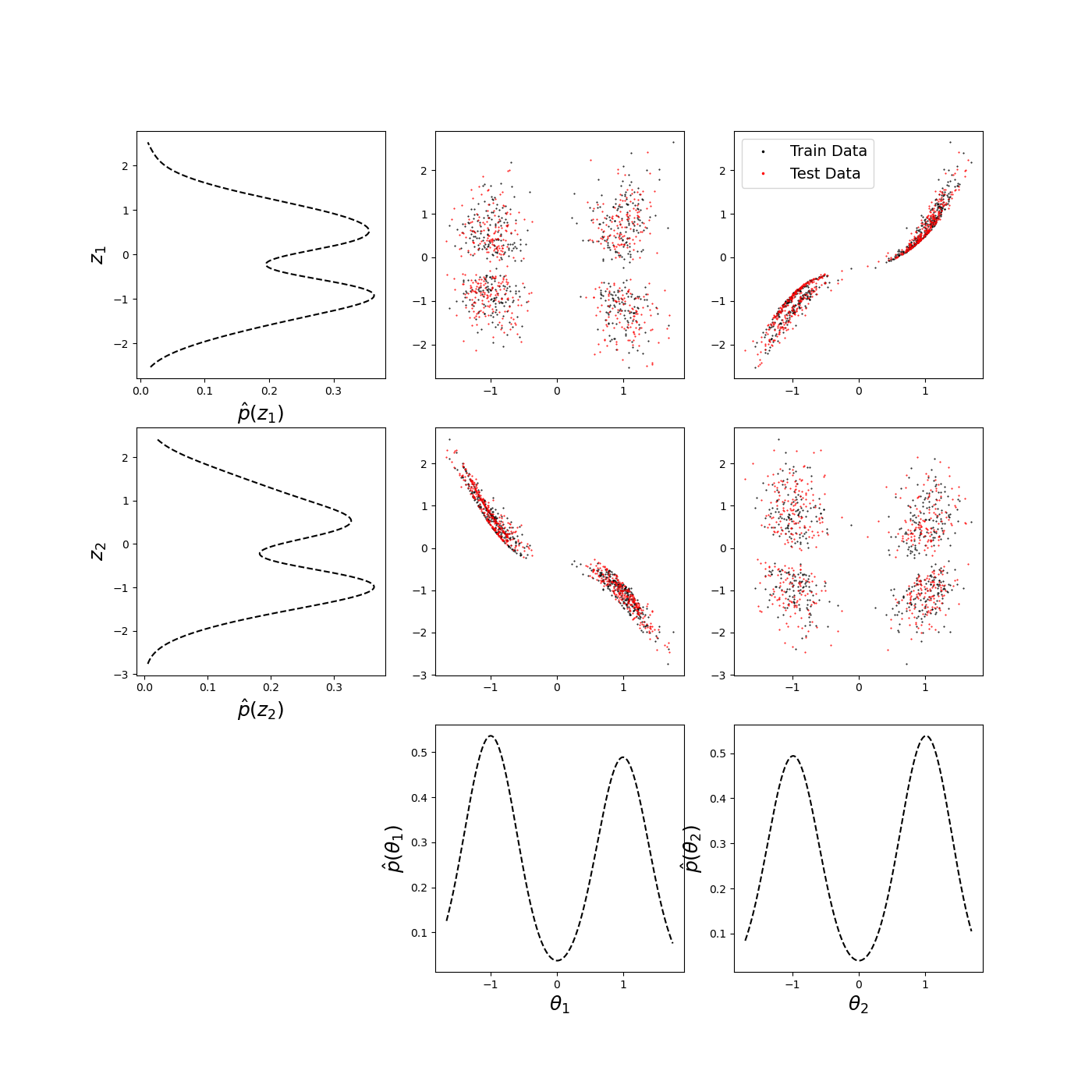}
    \includegraphics[width=.49\textwidth,angle=0,clip,trim=0pt 0pt 0pt 0pt]{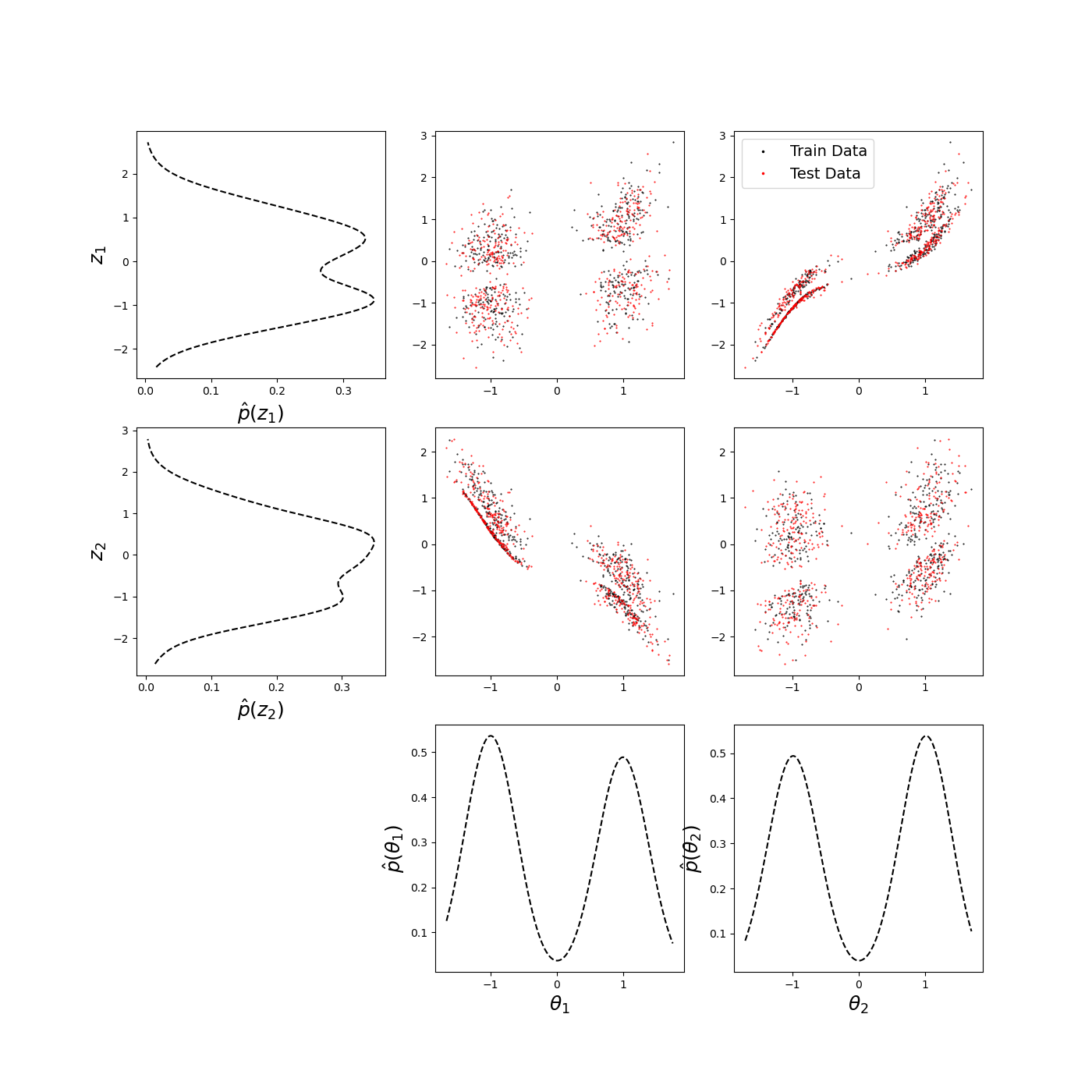}
    \caption{(\emph{top}) aggregated posterior comparison showing rotation of the latent space, (\emph{bottom}) worse disentanglement when latent space is rotated.}
    \label{fig:rot_dis}
\end{figure}

Training VAEs without hierarchical priors results in over-regularization more often than with the implementation of HP. Out of 50 trials, 10$\%$ trained without HP were unable to avoid over-regularization while all trials with HP successfully avoided over-regularization. More epochs are required in the reconstruction only phase (with and without HP) to avoid over-regularization than in previous examples. Disentanglement was never observed without the use of hierarchical priors. This again is due to rotation of the latent space relative to the generative parameter distribution due to rotationally invariant priors. To illustrate this concept, Figure \ref{fig:rot_dis} illustrates the effects of rotation of the latent space on disentanglement. Clearly, rotation dramatically impacts disentanglement, and the standard normal prior does not enforce any particular rotation of the latent space. 

\begin{figure}[h!]
    \centering
    \captionsetup{width=.75\linewidth}
    \includegraphics[width=.49\textwidth,angle=0,clip,trim=0pt 0pt 0pt 0pt]{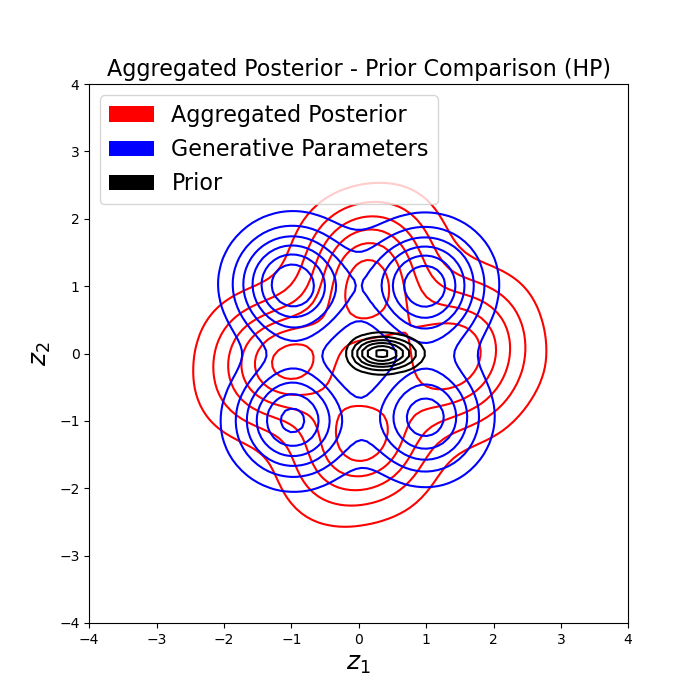}
    \caption{A 45 degree rotation of the latent space may be the result of local minima in the regularization loss during training.}
    \label{fig:rot_45}
\end{figure}

Implementing hierarchical priors, consistent observation of not only better reconstruction (avoiding over-regularization) but also reasonable disentanglement of the latent space in roughly half of all trained VAEs (out of 50) exemplifies the improved ability of hierarchical priors to produce a disentangled latent representation. Reconstruction of test samples is more accurate when implementing the hierarchical prior network, as illustrated in Figure \ref{fig:kle2_bimodal_gen_2_recon}. We hypothesize that disentanglement is observed in roughly half of our experiments due to local minima in the regularization loss corresponding to 45 degree rotations of the latent space (Figure \ref{fig:rot_45}). The learned priors using HP are often non-rotationally-invariant and aligned with the axes. However, the posterior is often rotated 45-degrees relative to this distribution, creating a non-factorized and therefore non-disentangled representation. 

\begin{figure}[h!]
    \centering
    \captionsetup{width=.75\linewidth}
    \includegraphics[width=.49\textwidth,angle=0,clip,trim=0pt 0pt 0pt 0pt]{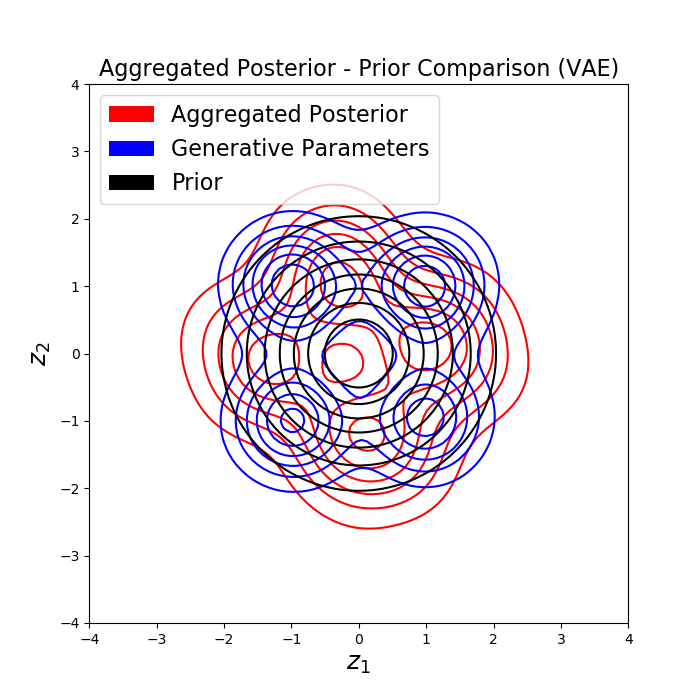}
    \includegraphics[width=.49\textwidth,angle=0,clip,trim=0pt 0pt 0pt 0pt]{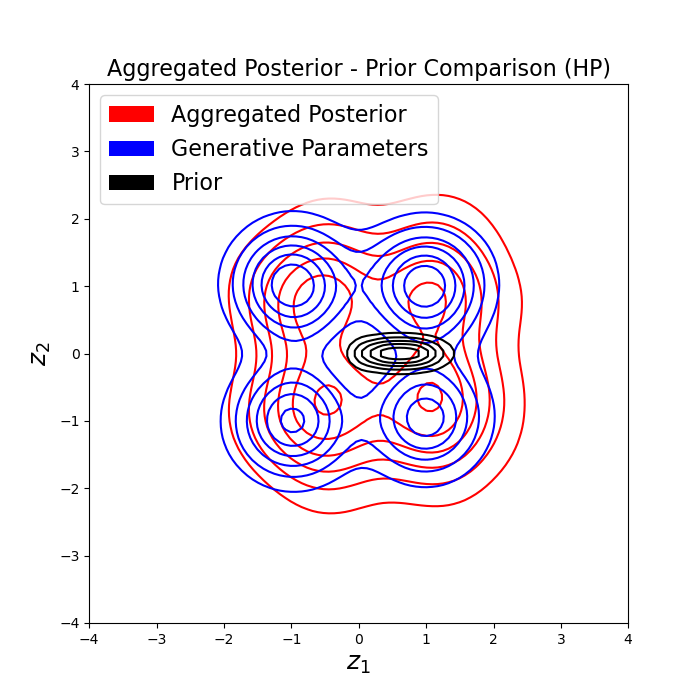}
    \caption{(\emph{left}) Aggregated posterior, prior, and generative parameter distribution comparison using VAE trained on multimodal generative parameter distribution without hierarchical network, (\emph{right}) with hierarchical network.}
    \label{fig:kle2_bimodal_gen_2_agg_post}
\end{figure}

Comparing $p(\theta)$, $p(z)$, and $q_\phi(z)$ with and without HP (Figure \ref{fig:kle2_bimodal_gen_2_agg_post}), stark differences are noticeable. Without the HP network, the aggregated posterior often captures the multimodality of the generative parameter distribution, but it is rotated relative to $p(\theta)$, creating a non-factorized $q_\phi(z)$. Training the VAE with hierarchical priors, the learned prior becomes non-rotationally invariant. The rotation of the aggregated posterior is therefore controlled by the orientation of the prior through the regularization loss, but mimics the shape of the generative parameter distribution. It is clear that the prior plays a significant role in terms of disentanglement: it controls the rotational orientation of the aggregated posterior.

\begin{figure}[h!]
    \centering
    \captionsetup{width=.75\linewidth}
    \includegraphics[width=.49\textwidth,angle=0,clip,trim=0pt 0pt 0pt 0pt]{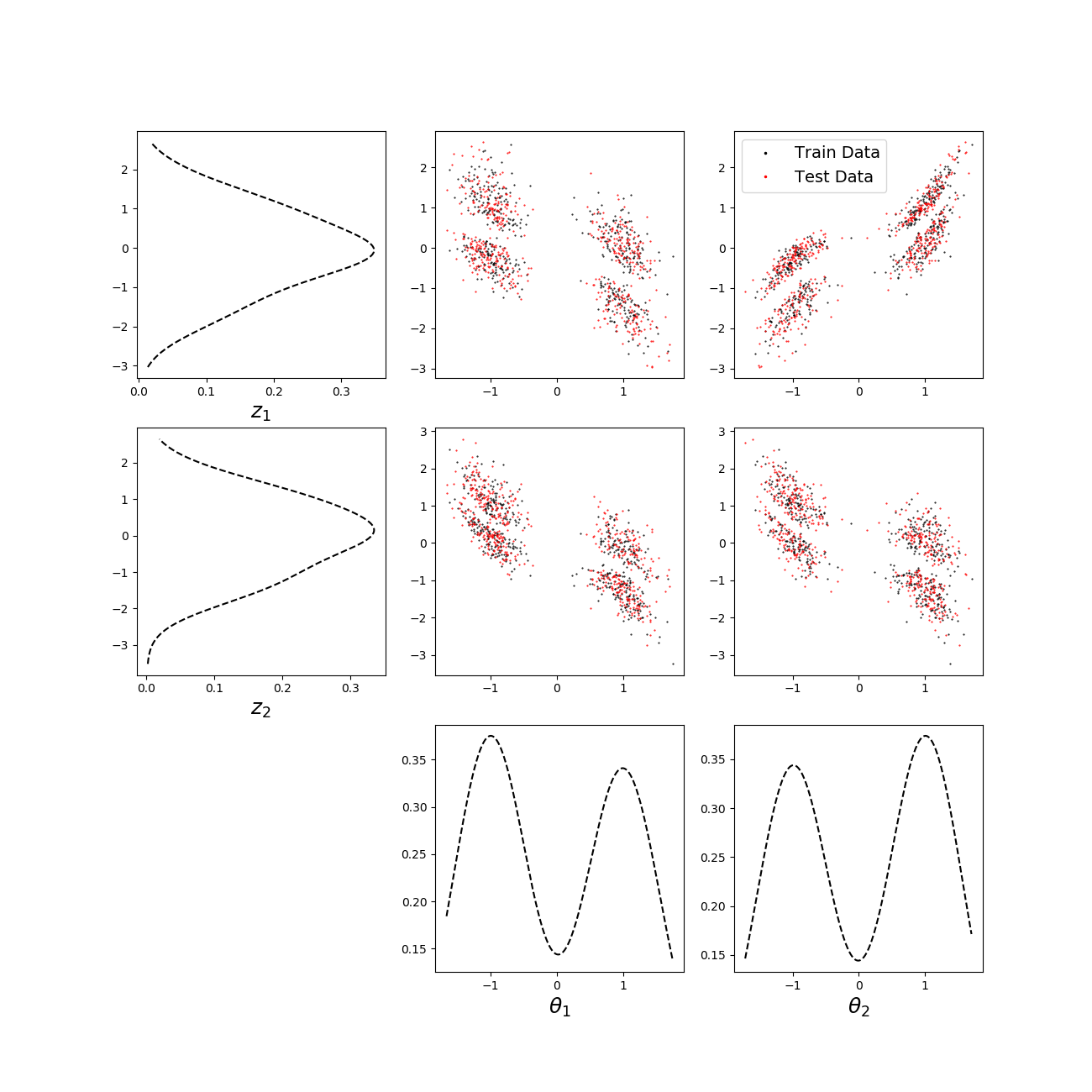}
    \includegraphics[width=.49\textwidth,angle=0,clip,trim=0pt 0pt 0pt 0pt]{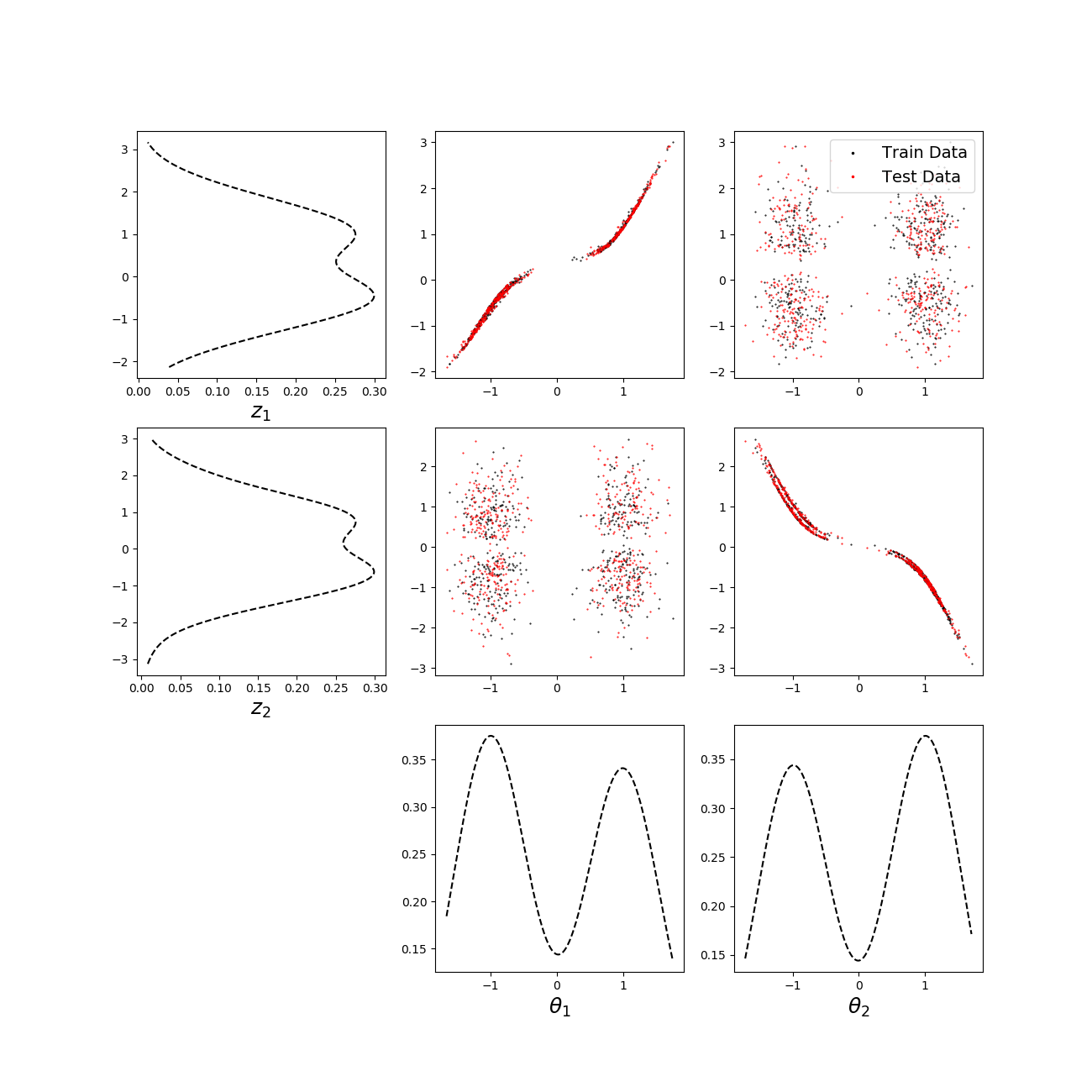}
    \caption{(\emph{left}) Qualitative disentanglement using VAE trained on multimodal generative parameter distribution without hierarchical network, (\emph{right}) with hierarchical network.}
    \label{fig:kle2_bimodal_gen_2_dis}
\end{figure}

A qualitative measure of disentanglement is compared in Figure \ref{fig:kle2_bimodal_gen_2_dis}. Without HP, the latent parameters are entangled; they are each weakly correlated  to both of the generative parameters. Adding HP to the VAE results in disentanglement in nearly half of our trials. When disentanglement does occur, each latent factor contains information on mostly a single but different generative factor. Through the course of our experiments, a relationship between disentanglement and the degree to which the aggregated posterior matches the generative parameter distribution is recognized. When disentanglement does not occur with the use of HP, the aggregated posterior is rotated relative to $p(\theta)$, or non-factorized (it has always been observed at around a 45-degree rotation). Only when $q_\phi(z)$ can be translated and scaled to better match $p(\theta)$, maintaining the correlations, does disentanglement occur. Thus, a quantitative measure of disentanglement (Eq \ref{eq:KL_dis}) is created from this idea. The KL divergence is estimated through sampling using the $k$-nearest neighbors ($k$-NN) approach (version $\epsilon 1$) found in \cite{4839047}. The optimization is performed using the  gradient-free Nelder-Mead optimization algorithm \cite{10.1093/comjnl/7.4.308}. 

\begin{figure}[h!]
    \centering
    \captionsetup{width=.75\linewidth}
    \includegraphics[width=.99\textwidth,angle=0,clip,trim=0pt 0pt 0pt 0pt]{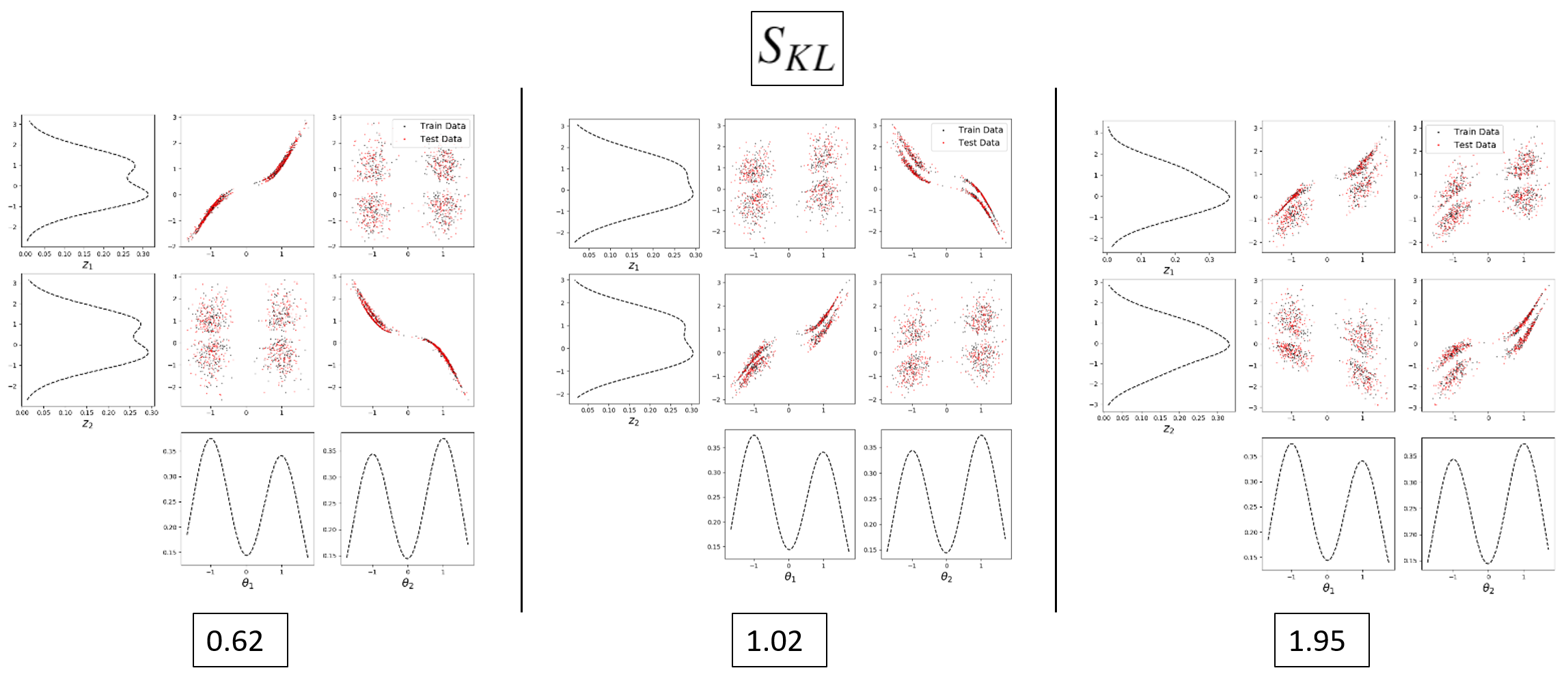}
    \caption{A quantitative measure of disentanglement compared to a qualitative measure. As $S_{KL}$ increases, the latent space becomes more entangled. }
    \label{fig:s_kl}
\end{figure}

In low-dimensional problems, humans are adept at determining disentanglement from qualitative measurements of disentanglement such as Figure  \ref{fig:kle2_bimodal_gen_2_dis}. It is, however, more difficult to obtain quantitative measurements of these properties. Figure \ref{fig:s_kl} shows the relationship between Eq. \ref{eq:KL_dis} and a qualitative measurement of disentanglement. Lower values of $S_{KL}$ indicate better disentanglement. This measure of disentanglement and the intuition behind it is discussed further in the conclusions section.

\section{Semi-Supervised Training}
Difficulties with consistently disentangling generative parameters have been illustrated up to this point with an unsupervised VAE framework. In some cases, however, generative parameters may be known for some number of samples, suggesting the possibility of a semi-supervised approach. These labeled samples can be leveraged to further improve the consistency of learning a disentangled representation. Consider data consisting of two partitions: labeled data $\{y^{(i)}, \theta^{(i)}\}_{i=1}^l$ and unlabeled data $\{y^{(i)}\}_{i=l+1}^{u+l}$. A one-to-one mapping between the generative parameters $\theta$ and the learned latent representation $z$ is sought when disentanglement is desired. Thus, enforcing the latent representation to match the generative parameters for labeled data in a semi-supervised approach should aid in achieving our desired objective more consistently. 

We begin the intuition behind a semi-supervised loss function by illustrating its connection to the standard ELBO VAE loss. One method of deriving the ELBO loss is to first expand the relative entropy between the data distribution and the induced data distribution to obtain
\begin{align*}\label{eq:elbo_exp}
    D_{KL}[p(y)||p_\psi(y)] & = -H(Y) + \mathbb{E}_{p(y)}[D_{KL}[q_\phi(z|y)||p(z)]] - \mathbb{E}_{p(y)}[D_{KL}[q_\phi(z|y)||p(z|y)]] \\
    & \;\;\;\;\;- \mathbb{E}_{p(y)q_\phi(z|y)}[\log p_\psi(y|z)]  
\end{align*}
where $-H(Y)$ is constant and the 'true' encoder $p(z|y)$ is unknown. Therefore, the term 
\[
\mathbb{E}_{p(y)}[D_{KL}[q_\phi(z|y)||p(z|y)]]
\] is usually ignored and we arrive at the ELBO, which upper bounds the left hand side. However, a relationship between $z$ and $y$ is known for labeled samples. This relationship can be used to assign $p(z^{(i)}|y^{(i)})$ on the labeled partition. For unlabeled data, the standard ELBO loss is still used for training and the semi-supervised loss to be minimized becomes
\begin{equation}\label{eq:ss_loss_pre}
    \mathcal{L}_{VAE-SS}(\phi,\psi) = \mathbb{E}_{p(y)}[D_{KL}[q_\phi(z|y)||p(z)]] - \mathbb{E}_{p_l(y)}[D_{KL}[q_\phi(z|y)||p(z|y)]] - \mathbb{E}_{p(y)q_\phi(z|y)}[\log p_\psi(y|z)] 
\end{equation}
where $p_l(y)$ is the distribution of inputs with corresponding labels, $p(y)$ is the distribution of all inputs (labeled and unlabeled), and $\mathbb{E}_{p_l(y)}[D_{KL}[q_\phi(z|y)||p(z|y)]]$ is denoted $\mathcal{L}_{SS}$.

Note that the term $\mathbb{E}_{p(y)}[D_{KL}[q_\phi(z|y)||p(z)] - D_{KL}[q_\phi(z|y)||p(z|y)]]$ is minimized by $q_\phi(z|y) = p(z|y)$ at $I(Z;Y)$, the mutual information between the generative parameters and high dimensional data. In unsupervised VAEs, the regularization term
$\mathbb{E}_{p(y)}[D_{KL}[q_\phi(z|y)||p(y)]]$ is minimized when $q_\phi(z|y) = p(y)$. As observed in previous sections, disentanglement is observed when the aggregated posterior is 'close' to the generative parameter distribution. With the semi-supervised loss being minimized when they are equivalent, the learned latent representations should be more easily and consistently disentangled. 

However, empirically it is found that this loss is very sensitive to changes in network parameters and unreasonably small learning rates are required for stability. Additionally, there is no obvious way to determine the variance of $p(z^{(i)}|y^{(i)})$ for each sample, only the mean is easily identifiable. We therefore propose to train with $\mathcal{L}_{SS} = \mathbb{E}_{p_l(y)}[-\log q_\phi(z|y)]$ instead such that the loss function becomes 
\begin{equation}\label{eq:ss_loss}
    \mathcal{L}_{VAE-SS}(\phi,\psi) = \mathbb{E}_{p(y)}[D_{KL}[q_\phi(z|y)||p(z)]] - \mathbb{E}_{p_l(y)}[\log q_\phi(z|y)] - \mathbb{E}_{p(y)q_\phi(z|y)}[\log p_\psi(y|z)] \;. 
\end{equation}
Training with this loss achieves the desired outcome of consistently learning disentangled representations while being simple and efficient to implement. 

\begin{figure}[h!]
    \centering
    \captionsetup{width=.75\linewidth}
    \includegraphics[width=.49\textwidth,angle=0,clip,trim=0pt 0pt 0pt 0pt]{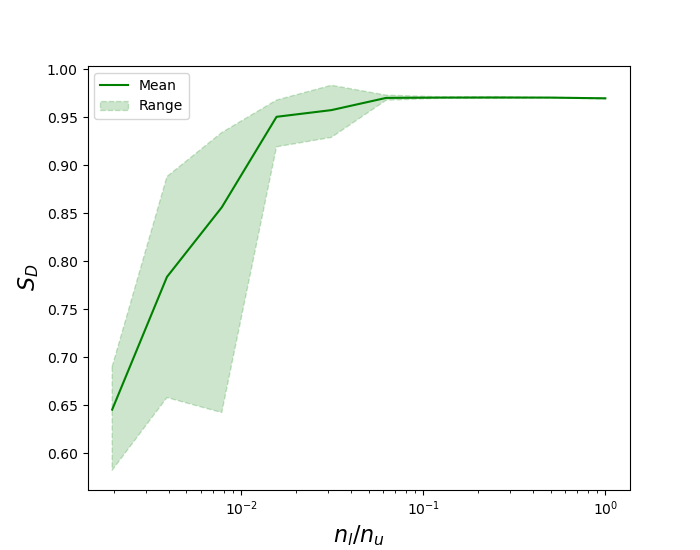}
    \includegraphics[width=.49\textwidth,angle=0,clip,trim=0pt 0pt 0pt 0pt]{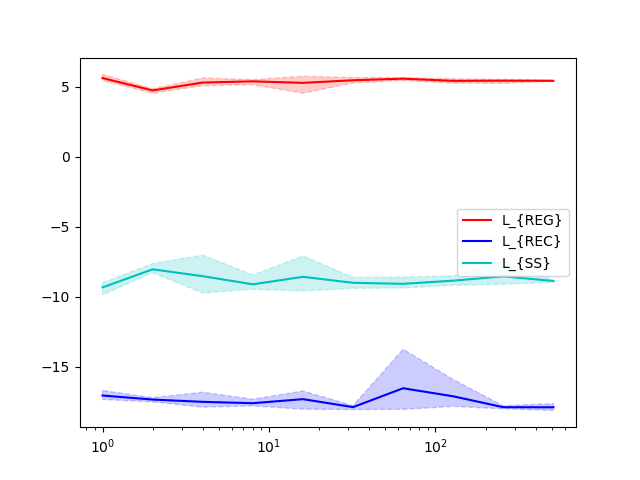}
    \caption{\emph{(left)} Disentanglement score mean increases with ratio of labeled to unlabeled samples when training with a semi-supervised loss. Disentanglement also becomes more consistently observed. \emph{(right)} Training losses are unaffected by the number of labeled samples. }
    \label{fig:ss_dis}
\end{figure}

Incorporating some labeled samples into training the VAE, a disentangled latent representation can be consistently learned. Figure \ref{fig:ss_dis} illustrates the relationship between increasing the number of labeled samples and the disentanglement score of the learned latent representation. In each case, there are 512 unlabeled samples. Each trial varies in the number of labeled samples, and VAEs trained with the same number of labeled samples are trained with a different set of labeled samples.

The training losses do not seem to be effected by the number of labeled samples, only the disentanglement score is effected. With a low number of labeled samples, the semi-supervised VAE trains very similarly to the unsupervised VAE. That is, disentanglement is observed rather randomly, and the learned latent representation varies dramatically between trials. Labeling around $1\%$ of the samples begins to result in consistently good disentanglement. Labeling between $3\%$ and $8\%$ results in learning disentangled latent representations which are nearly identical between trials. It follows from these results that disentangled representations can be consistently learned when training with Eq. \ref{eq:ss_loss} when using a sufficient number of labeled samples (assuming a sufficiently expressive architecture).

\begin{figure}[h!]
    \centering
    \captionsetup{width=.75\linewidth}
    \includegraphics[width=.49\textwidth,angle=0,clip,trim=0pt 0pt 0pt 0pt]{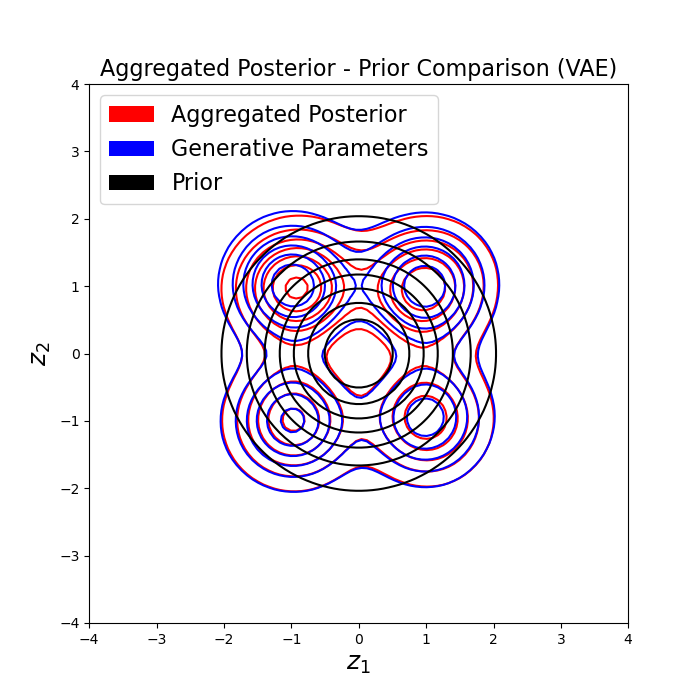}
    \includegraphics[width=.49\textwidth,angle=0,clip,trim=0pt 0pt 0pt 0pt]{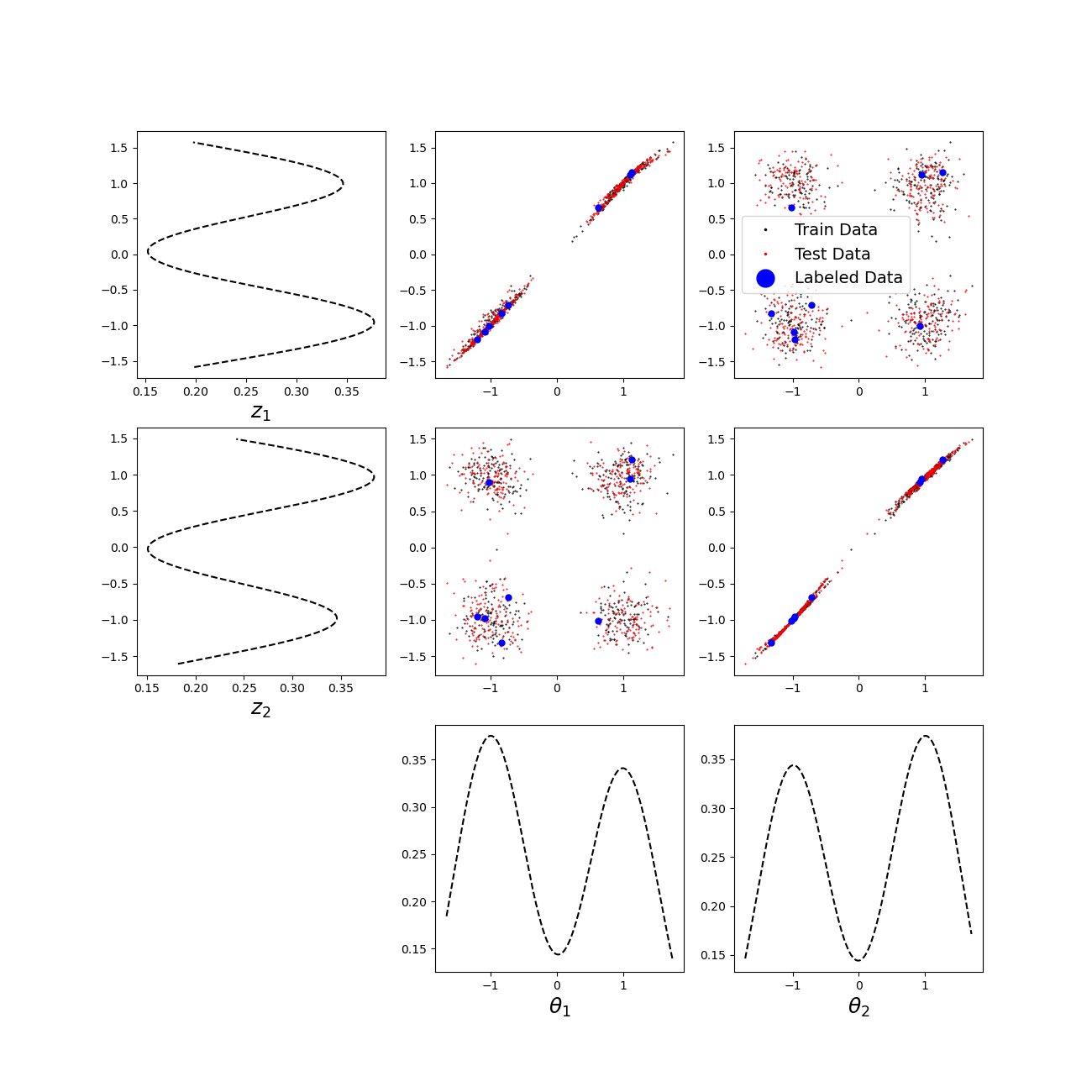}
    \caption{\emph{(left)} Aggregated posterior matches the generative parameter distribution with semi-supervised training. \emph{(right)} Multi-modality is well preserved. }
    \label{fig:agg_post_ss}
\end{figure}

Using a semi-supervised method also improves the ability of the VAE to predict data in regions of lower density. In Figure \ref{fig:agg_post_ss}, we observe that the aggregated posterior matches the generative parameter distribution much better than the unsupervised case with just over $1\%$ of the samples labeled. Additionally, regions of low density in the generative parameter distribution are better represented in the semi-supervised case over the unsupervised case; in other words, multimodality is better preserved (compare to Figures \ref{fig:kle2_bimodal_gen_2_agg_post} and \ref{fig:kle2_bimodal_gen_2_dis}).

\section{Concluding Remarks and Perspectives}\label{sec:conclusion}
Learning representations such that each latent dimension corresponds to a single physical generative factor of variation is useful in many applications, particularly when learned in an unsupervised manner. 
Learning such {\em disentangled} representations using VAEs is dependent on many factors including network architecture, assumed form of distributions, prior selection, hyperparameters, and random seeding. 
The goal of our work is to develop (i) a  consistent unsupervised framework to learn disentangled representations of data obtained through physical experiments or PDE simulations, and (ii) to comprehensively characterize the underlying training process, and to recommend strategies to avoid sub-optimal representations. 

Accurate reconstruction is desirable from a variety of perspectives, including being necessary for consistent disentanglement. Given two samples near one another in data space, and an accurate decoder, those two samples will be encourage to be near one another in latent space. This is a result of the sampling operation when computing the reconstruction loss. The reconstruction loss is minimized if samples near one another in latent space correspond to samples near one another in data space. Thus, finding an architecture suitable for accurate prediction from latent representations is of great importance to learn disentangled representations. In our experiments, different architectures were implemented before arriving at and refining the dense architecture (Section \ref{sec:architecture}), which was found to accurately reconstruct the data from latent codes. Even with a suitable architecture, however,  significant obstacles need to be overcome to arrive at a consistent framework for achieving disentangled representations. Over-regularization can often be difficult to avoid, especially when the variation among data samples is minor. This again emphasizes the necessity to accurately reconstruct the data first before attempting to learn meaningful representations. We have illustrated methods of avoiding over-regularization when training VAEs, but rotationally invariant priors can still create additional difficulties in the ability to disentangle parameters. We illustrated in Section \ref{sec:KLE2} that the standard normal prior typically assumed (which is rotationally-invariant) does not enforce any particular rotation of the latent space, often leading to entangled representations. Rotation of the latent space matters greatly, and without rotational enforcement on the encoder, disentanglement is rarely, or rather randomly, achieved when training with the ELBO loss. We have also shown that the implementation of hierarchical priors allows one to learn non-rotationally-invariant priors such that the regularization loss enforces a rotational constraint on the encoding distribution. However, the regularization loss can contain local minima as the latent space rotates, enforcing a non-factorized and thus incorrectly rotated aggregated posterior. This indicates the need for better prior selection, especially in higher latent dimensions when rotations create more complex effects (\ref{app:n10}). 

Matching the aggregated posterior to the generative parameter distribution can also be enforce by including labeled samples during training. Including some number of labeled samples in the dataset and training with a semi-supervised loss, the aggregated posterior consistently matches the shape and orientation of the generative parameter distribution, effectively learning a disentangled representation. The multimodality of the data distribution is also better represented when using labeled data, indicating that the VAE can better predict data in regions of low density over the unsupervised version. 

In reference to Section \ref{sec:KLE2}, the total correlation (TC)  $D_{KL}[q_\phi(z)||\prod_{i=1}^n q_\phi(z_i)]$ appears to be a useful and simpler measurement of disentanglement. When the generative parameters are completely independent (i.e. $p(\theta) = \prod_{i=1}^p p(\theta_i)$) and disentanglement occurs when a factorized $q_\phi(z)$ is learned (aligning the latent space axes with the generative parameter axes). This is the objective of the FactorVAE framework \cite{kim2019disentangling}, which can successfully encourage a factorized $q_\phi(z)$ through the introduction of TC into the loss function, modifying to ELBO. However, considering a case in which the generative parameters are correlated, a factorized $q_\phi(z)$ is not necessarily desirable. It is in anticipation of a more correlated $p(\theta)$ that we use Eq \ref{eq:KL_dis} as a measure of disentanglement. Additionally, in our work we do not modify the standard VAE objective to produce more accurate reconstruction of the data. 

\begin{figure}[h!]
    \centering
    \captionsetup{width=.75\linewidth}
    \includegraphics[width=.49\textwidth,angle=0,clip,trim=0pt 0pt 0pt 0pt]{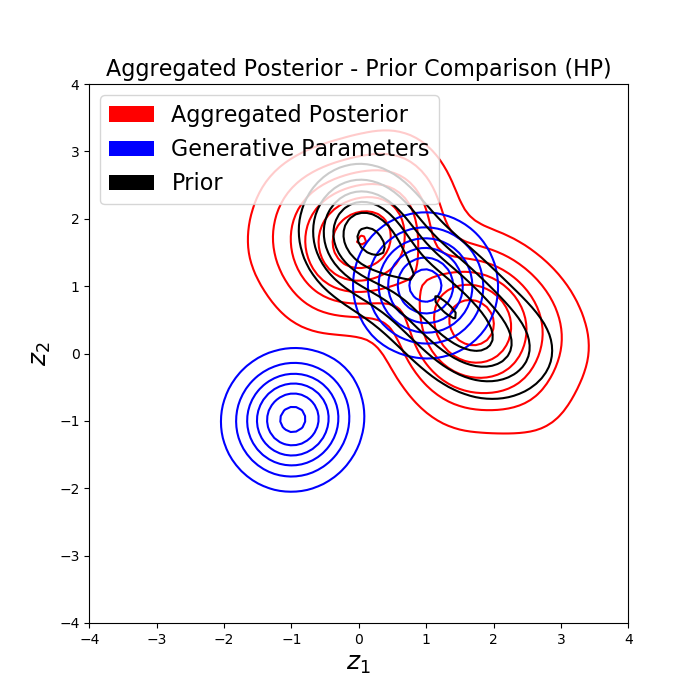}
    \includegraphics[width=.49\textwidth,angle=0,clip,trim=0pt 0pt 0pt 0pt]{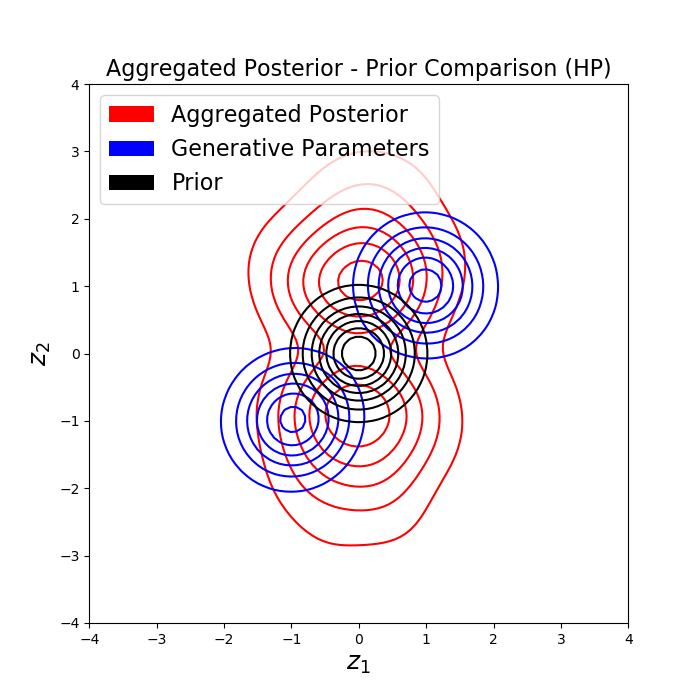}
    \includegraphics[width=.49\textwidth,angle=0,clip,trim=0pt 0pt 0pt 0pt]{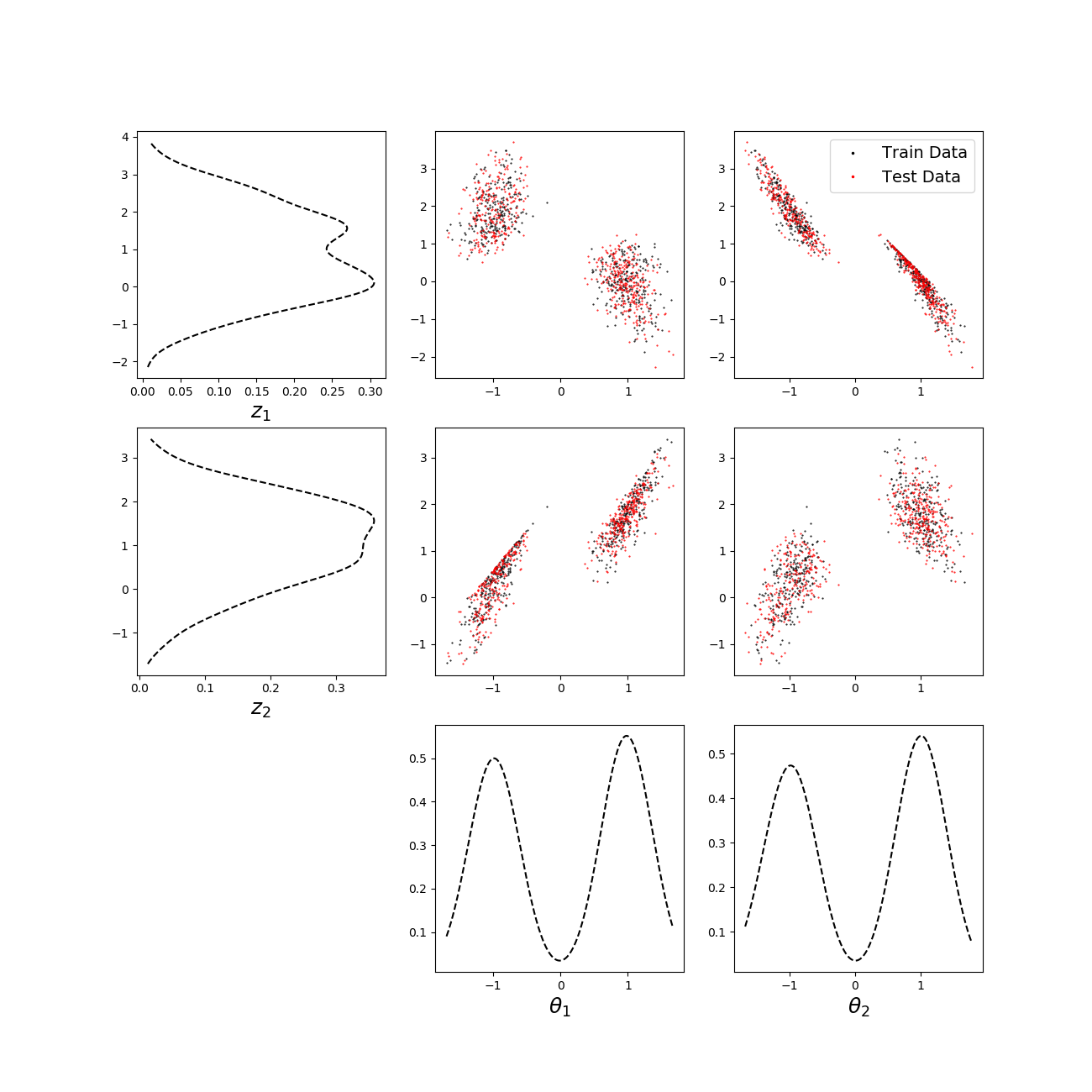}
    \includegraphics[width=.49\textwidth,angle=0,clip,trim=0pt 0pt 0pt 0pt]{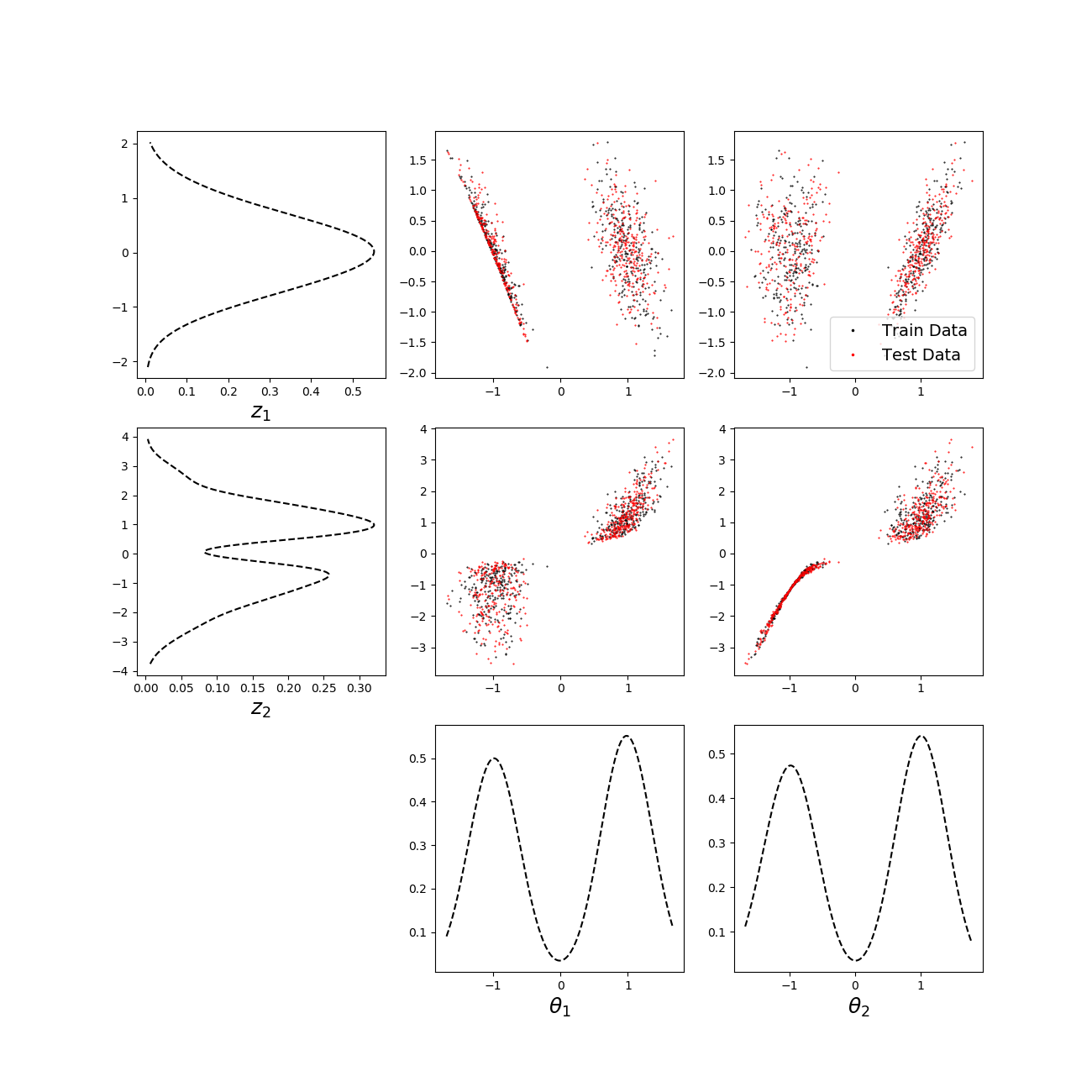}
    \caption{(\emph{top}) aggregated posterior comparison correlations / rotations relative to the generative parameter distribution, (\emph{bottom}) worse disentanglement when correlations not expressed in latent space.}
    \label{fig:corr_case}
\end{figure}

Complete disentanglement has not been observed when generative parameters are correlated, but after many trials the same conclusions have been drawn as the uncorrelated case: for disentanglement to occur, the aggregated posterior must contain the same 'shape' as the generative parameter distribution - this includes correlations up to permutations of the axes. Figure \ref{fig:corr_case} illustrates these ideas, though we note not perfectly. Future work will include disentangling correlated generative parameters, which may be facilitated through learning correlated priors using HP.

In addition to disentangling correlated generative parameters, our broader aim is to extend our work to more complex problems to create a general framework for consistent unsupervised or semi-supervised representation learning. Through our observations here regarding non-rotationally invariant priors along with insights gained from \cite{rolinek2019variational}, we hypothesize that such a framework will be largely focused on both prior selection and the structural form of the encoding and decoding distributions. Additionally, in a completely unsupervised setting, one must find an encoder and decoder which disentangle the generative parameters, but the dimension of the generative parameters may be unknown. The dimension of the latent space is always user-specified; if the dimension of the latent space is too small or too large, how does this effect the learned representation? Can one successfully and consistently disentangle generative parameters in higher dimensions? These are some of the open  questions to be addressed in the future.

The issue of over-regularization often greatly hinders our ability to train VAEs (Section \ref{sec:over_reg}). Different initialization strategies may be investigated to increase training performance and avoid the issue altogether. It has been shown that principled selection of activation functions, architecture, and initialization can greatly improve not only the efficiency of training, but also facilitate greater performance in terms of reconstruction \cite{sitzmann2020implicit}. 

The greater scope of this work is to develop an unsupervised and interpretable representation learning framework to generate probabilistic reduced order models for physical problems and use learned representations for efficient design optimization.

\section*{Acknowledgments}
{The authors acknowledge  support from the Air Force under the grant FA9550-17-1-0195 (Program Managers: Dr. Mitat Birkan and Dr. Fariba Fahroo).
Computing resources were provided by the NSF via grant 1531752 (Program Manager: Dr. Stefan Robila).
}
\bibliography{ref.bib}

\appendix

\section{Rotationally-Invariant Distributions} \label{app:rot_inv}
A matrix $\mathbf{R} \in \mathbb{R}^{n\times n}$ is a rotation matrix if for all $z \in \mathbb{R}^n$, $\lVert \mathbf{R}z \rVert_2 = \lVert z \rVert_2$. 

A probability distribution $p(z)$ is said to be \emph{rotationally-invariant} if $p(z) = p(\mathbf{R}z)$ for all $z \in \mathbb{R}^n$ and for all rotation matrices $\mathbf{R} \in \mathbb{R}^{n\times n}$.

The ELBO loss (Eq. \ref{eq:L_VAE}) is unaffected by rotations of the latent space when training with a rotationally-invariant prior. This is shown in detail in Ref. \cite{rolinek2019variational}. 

\section{Higher Latent Dimensions}\label{app:n10}
As discussed in Sections \ref{sec:beta_VAE}, \ref{sec:KLE2}, and \ref{sec:conclusion}, rotation of the latent space greatly impacts disentanglement. In a two dimensional latent space ($n=2$) with rotationally invariant prior, it is possible to converge to a disentangled latent space due to favorable random initialization or only slightly encourage rotational constraints on the latent space (factorization). In higher dimensions, however, many more axes of rotation exist. Thus, the probability of converging to a latent space which is rotated adequately for disentanglement to occur is far lower. In this case, some method of enforcing rotation of the latent space must be considered. In our experiments, we observed that higher dimensional latent spaces introduce over-regularization of the prior encoder and prior decoder with the implementation of hierarchical priors. Thus, there are many more challenges to overcome toward the path of disentanglement with increased latent dimension without spoiling the ELBO. A scheduler similar to the $\beta$ scheduler described in Section \ref{sec:issues} is implemented to prevent over-regularization of the sub-latent space. 

\begin{comment}
Figure \ref{fig:n10} illustrates qualitative disentanglement when implementing a classic VAE on a KLE10 dataset with $p(\theta)$ the standard normal distribution. Twenty VAEs were trained both with and without hierarchical priors, and similar results were illustrated in each case. 

\begin{figure}[h!]
    \centering
    \captionsetup{width=.75\linewidth}
    \includegraphics[width=.9\textwidth,angle=0,clip,trim=0pt 0pt 0pt 0pt]{./Figures/dis_hp_0.PNG}
    \caption{Higher dimensional latent spaces ($n=10$ here) are more difficult to disentangle.}
    \label{fig:n10}
\end{figure}
\end{comment}

\begin{figure}[h!]
    \centering
    \captionsetup{width=.75\linewidth}
    \includegraphics[width=.49\textwidth,angle=0,clip,trim=0pt 0pt 0pt 0pt]{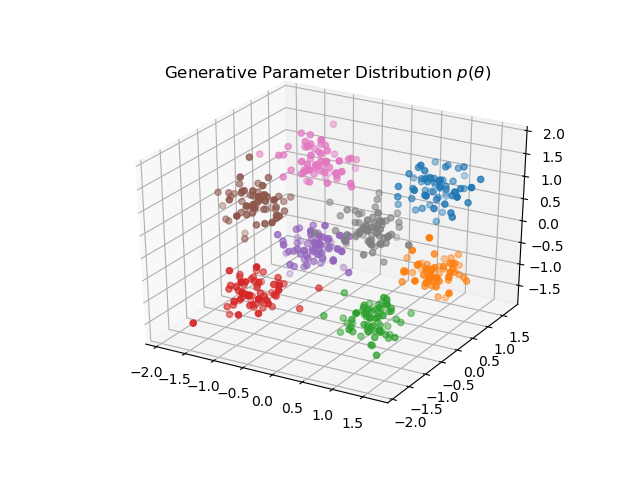}
    \includegraphics[width=.49\textwidth,angle=0,clip,trim=0pt 0pt 0pt 0pt]{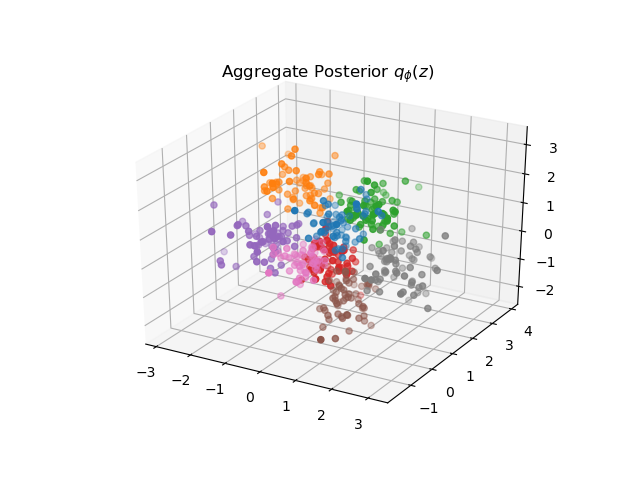}
    \caption{Comparison between (\emph{left}) the generative parameter distribution and (\emph{right}) the aggregated posterior with $p, \; n = 3$. The aggregated posterior matches the shape of the generative parameter distribution, but is not fully factorized.}
    \label{fig:n3}
\end{figure}

To understand some of the problems introduced by increasing the latent dimension, VAEs with HP were trained on the KLE3 dataset with a multimodal $p(\theta)$. In most cases (12 out of 20 trials), the learned aggregated posterior matched the shape of the generative parameter distribution, but was relatively rotated (thus not factorized). The other 8 of 20 trials illustrated aggregated posteriors which did not match the generative parameter distribution in shape. Figure \ref{fig:n3} shows a 3D comparison in the aggregated posterior and generative parameter distribution, showing weak disentanglement properties due to incorrect rotations. Incorrect rotation of the latent space may be caused by local minima in the regularization loss. Non-rotationally-invariant priors are learned with HP, but still the latent space is rotated incorrectly and does not become factorized. 

Another 15 trials were performed using the VAE without HP, but with a non-rotationally-invariant Gaussian prior rather than the rotationally-invariant standard normal. Similar results were observed as with HP. The multimodal properties of the generative parameter distribution are sufficiently captured by the aggregated posterior, but the correlations are not captured. In all cases mentioned, with or without HP, reconstruction accuracy is high. 


\end{document}